\documentclass[showpacs,preprintnumbers,amsmath,amssymb,APSl,prd,nofootinbib,superscriptaddress, twocolumn]{revtex4-1}

\usepackage{dcolumn}
\usepackage{bm}
\usepackage{ifpdf}
\usepackage{hyperref}
\usepackage{bm}
\usepackage{xcolor,color,graphicx,graphics}
\usepackage[spanish,english]{babel}
\usepackage[latin1]{inputenc}
\usepackage[OT1]{fontenc}
\usepackage{latexsym,amssymb,amsmath,amsfonts}
\usepackage{makeidx}
\usepackage{graphicx}
\usepackage{epsfig}
\usepackage{natbib}
\usepackage{epstopdf}
\usepackage{mathrsfs}
\usepackage{hyperref}
\hypersetup{colorlinks=true, linkcolor=blue, citecolor=green}
\usepackage{enumerate}
\usepackage{verbatim} 
\usepackage{ulem}

\newcommand{\R}{\mathcal{R}}

\begin{document}
\title{Numerical evolutions of boson stars in Palatini $f(\R)$ gravity}
\author{Andreu Mas\'o-Ferrando} \email{andreu.maso@uv.es}
\affiliation{Departamento de F\'{i}sica Te\'{o}rica and IFIC, Centro Mixto Universidad de Valencia - CSIC.
    Universidad de Valencia, Burjassot-46100, Valencia, Spain}
\author{Nicolas Sanchis-Gual} \email{nicolas.sanchis@uv.es}
\affiliation{Departamento de Astronom\'{\i}a y Astrof\'{\i}sica, Universitat de Val\`encia,
    Dr. Moliner 50, 46100, Burjassot (Val\`encia), Spain}	
    \affiliation{Departamento  de  Matem\'{a}tica  da  Universidade  de  Aveiro  and  Centre  for  Research  and  Development in  Mathematics  and  Applications  (CIDMA),  Campus  de  Santiago,  3810-183  Aveiro,  Portugal} 
\author{Jos\'e A. Font} \email{j.antonio.font@uv.es}
\affiliation{Departamento de Astronom\'{\i}a y Astrof\'{\i}sica, Universitat de Val\`encia,
    Dr. Moliner 50, 46100, Burjassot (Val\`encia), Spain}
\affiliation{Observatori Astron\`omic, Universitat de Val\`encia, Catedr\'atico 
    Jos\'e Beltr\'an 2, 46980, Paterna (Val\`encia), Spain}
\author{Gonzalo J. Olmo} \email{gonzalo.olmo@uv.es}
\affiliation{Departamento de F\'{i}sica Te\'{o}rica and IFIC, Centro Mixto Universidad de Valencia - CSIC.
    Universidad de Valencia, Burjassot-46100, Valencia, Spain}
\affiliation{Universidade Federal do Cear\'a (UFC), Departamento de F\'isica,\\ Campus do Pici, Fortaleza - CE, C.P. 6030, 60455-760 - Brazil.}

\date{\today}
\begin{abstract}
We investigate the time evolution of spherically symmetric boson stars in Palatini $f(\R)$ gravity through Numerical Relativity computations. Employing a novel approach that establishes a correspondence between modified gravity with scalar matter and General Relativity with modified scalar matter, we are able to use the techniques of Numerical Relativity to simulate these systems. Specifically, we focus on the quadratic theory $f(\R)=\R+\xi\R^2$ and compare the obtained solutions with those in General Relativity, exploring both positive and negative values of the coupling parameter $\xi$. Our findings reveal that boson stars in Palatini $f(\R)$ gravity exhibit both stable and unstable evolutions. The latter give rise to three distinct scenarios: migration towards a stable configuration, complete dispersion, and gravitational collapse leading to the formation of a baby universe structure.
\end{abstract}

\maketitle

\section{Introduction}
Boson stars~\cite{Liebling:2012fv} are gravitationally bound configurations of bosonic particles that are minimally coupled to gravity. Their constituent particle is described by a massive oscillating complex scalar (or vector) field, whose dispersive nature balances the gravitational pull generated by itself. Boson stars masses and sizes range from atomic to astrophysical scales, depending on the mass of the bosonic particle. The study of boson stars has a rich history, starting from the groundbreaking work by Kaup~\cite{Kaup:1968zz} in 1968 and Ruffini and Bonazzola~\cite{Ruffini:1969qy} in 1969. Proca stars, the vector boson star counterparts, were more recently proposed~\cite{Brito:2015pxa}. Since then, boson stars have remained a subject of intensive investigation in various theoretical frameworks, and their properties and dynamics have been the focus of active research in astrophysics and cosmology (see~\cite{Schunck:2003kk,Liebling:2012fv,CalderonBustillo:2020fyi}).

In recent years, significant progress has been made in the development of numerical codes for computing the properties of boson stars, employing hyperbolic formulations of Einstein's equations, along with appropriate gauge conditions. These codes have been used to study various aspects of boson stars, including their stability~\cite{Gleiser:1988ih,Guzman:2009xre,Sanchis-Gual:2019ljs}, their formation through gravitational collapse of a dilute cloud of bosonic particles~\cite{Seidel:1993zk, DiGiovanni:2018bvo}, and the presence of scalar remnants around black holes~\cite{Sanchis-Gual:2014ewa,Escorihuela-Tomas:2017uac}. Additionally, the study of boson stars has also expanded to include modified gravity theories, such as Palatini $f(\R)$ gravity~\cite{Maso-Ferrando:2021ngp, Maso-Ferrando:2023nju}, which provides an alternative description of gravity in the strong field regime.

Boson stars, which have the capability to achieve higher densities compared to other astrophysical compact objects \cite{Olmo:2019flu}, offer a promising avenue for investigating potential modifications to the gravitational sector in the regime of strong gravitational fields. Unlike black holes, boson stars lack a horizon, which implies that the innermost regions of these objects could potentially be observed, offering new insights into the extension of Einstein's gravity. In this regard, $f(\R)$ theories~\cite{DeFelice:2010aj,Olmo:2011uz} provide a convenient framework for studying the properties and dynamics of boson stars beyond the standard model set by General Relativity (GR), since they come with a great amount of freedom while keeping the field equations within reasonable limits of simplicity.

Traditionally, space-time has been assumed to be described by a Riemannian geometry solely determined by the metric tensor. However, an alternative approach, known as the metric-affine or Palatini approach~\cite{Olmo:2011uz,BeltranJimenez:2019esp,Hehl:1994ue}, considers that the metric and the affine connection are independent. While this choice has no impact in the derivation of the field equations of GR  minimally coupled to scalar fields, it becomes relevant when considering $f(\R)$ gravity (and non-minimally coupled fields). The Palatini approach to $f(\R)$ gravity provides a suitable framework for testing the strong-field regime of gravitational interactions without encountering conflicts with current solar system observations or gravitational-wave astronomy results~\cite{Baker:2017hug,Creminelli:2017sry,Ezquiaga:2017ekz,Ezquiaga:2018btd,Sakstein:2017xjx,Lombriser:2015sxa,Olmo:2005zr}. 

Moreover, there exists a correspondence between the space of solutions of GR and Ricci-based gravity theories, a family of models in which $f(\R)$ gravity is included \cite{Olmo:2022rhf}. This opens up the possibility of using techniques developed for GR, such as Numerical Relativity, for solving problems in modified gravity scenarios~\cite{Afonso:2018bpv}. This correspondence allows us to compute the solutions of a canonical boson star in $f(\R)$ gravity  by considering the alternative problem of a non-linear (or non-canonical) complex scalar field matter Lagrangian coupled to GR~\cite{Maso-Ferrando:2021ngp}. 

In this work, we aim to investigate the time evolution of boson stars in Palatini $f(\R)$ gravity using state-of-the-art numerical techniques. By studying boson stars in Palatini $f(\R)$ gravity, we seek to understand the effects of modified gravity on the properties and dynamics of these compact objects. The findings reported in this paper may shed light on the fundamental nature of gravity in the strong gravitational regime and contribute to our understanding of the astrophysical implications of modified gravity theories. The interested reader is addressed to~\cite{Maso-Ferrando:2021ngp, Maso-Ferrando:2023nju} where our earlier results on this line of research were presented. 

This paper is organized as follows: In Section~\ref{sec:framework} we present the correspondence between $f(\R)$ gravity and GR and the time evolution formalism. Section~\ref{sec:inidata} discusses the initial data we build in order to perform the time evolutions. The numerical method used in the simulations is described in Section \ref{sec:numerical}. Our results are presented and discussed in Section \ref{sec:results}. Finally, we end up summarizing our findings and discussing our future perspectives in Section \ref{sec:final}. Tests of the code are reported in Appendix \ref{sec:convergence}. Those involve a convergence analysis under grid refinement and a monitoring of the numerical violations of the constraint equations. We use Greek indices $\alpha, \beta, ...$ when referring to spacetime indices, while Latin indices $i, j, ...$ are used for spatial indices. Moreover, we adopt geometrized units $c=G=1$ throughout this work.

\section{Framework}\label{sec:framework}

\subsection{Theory correspondence}

The action of a  boson star in Palatini $f(\R)$ takes the form 
\begin{equation}\label{eq:Ac1}
    S_{f(\mathcal{R})}=\int d^4x \sqrt{-g} \frac{f(\mathcal{R})}{2 \kappa} -\frac{1}{2}\int d^4x \sqrt{-g} P(X,\Phi) \quad .
\end{equation}
where gravity is described in terms of a Palatini $f(\mathcal{R})$ function and the matter sector is represented by a complex scalar field $\Phi$ with canonical Lagrangian
\begin{equation}
    P(X,\Phi)=X-2V(\Phi)\quad,
\end{equation} 
where $X=g^{\alpha \beta}\partial_{\alpha}\Phi^*\partial_{\beta}\Phi$, $V(\Phi)=-\mu^2 \Phi^* \Phi/2$,  the scalar field mass is $\mu$, $g=\det (g_{\alpha\beta})$, and $\kappa=8\pi$. Here, we are defining $\mathcal{R}=g^{\mu\nu}R_{\mu\nu}(\Gamma)$, with $R_{\mu\nu}(\Gamma)$ representing the Ricci tensor of a connection $\Gamma^\lambda_{\alpha\beta}$ a priori independent of the metric $g_{\mu\nu}$. 

Manipulating the field equations that follow from independent variations of the metric and the connection, one finds that by introducing an auxiliary metric
\begin{equation}\label{eq:conformal}
    q_{\mu\nu}\equiv f_{\mathcal{R}} g_{\mu\nu} \ ,
\end{equation}
the explicit relation between  $\Gamma^\lambda_{\alpha\beta}$ and $g_{\mu\nu}$ is determined by 
\begin{equation}
    \Gamma^\lambda_{\mu\nu}=\frac{q^{\lambda\rho}}{2}\left[\partial_\mu q_{\rho\nu}+\partial_\nu q_{\rho\mu}-\partial_\rho q_{\mu\nu}\right] \ ,
\end{equation}
with $f_\mathcal{R}\equiv \partial f/\partial \mathcal{R}$. Then, the connection $\Gamma^\lambda_{\alpha\beta}$ is the Levi-Civita connection of the auxiliary metric $q_{\mu\nu}$. We note that the conformal factor $f_{\mathcal{R}}$ must be regarded as a function of the metric $g_{\alpha\beta}$ and the matter fields which is specified by the algebraic equation
\begin{equation}\label{eq:Trace}
    \mathcal{R}f_{\mathcal{R}}-2f=\kappa T \ ,
\end{equation}
where $T$ represents the trace of the stress-energy tensor, which is defined as
\begin{equation}\label{eq:Tmn-g}
    T_{\mu \nu}\equiv-\frac{2}{\sqrt{-g}}\frac{\delta(\sqrt{-g} P(X,\Phi))}{\delta g^{\mu \nu}} \ .
\end{equation}
For simplicity, and to make contact with the existing literature, we will specify the gravity Lagrangian by the quadratic function
\begin{equation}
    f(\mathcal{R})=\mathcal{R}+\xi \mathcal{R}^2 \ .
\end{equation}
This is the Palatini version of the so-called Starobinsky model \cite{Starobinsky:1980te}, and represents the $R-$dependent part of the quantum-corrected extension of GR when quantum matter fields are considered in a curved space-time. Within the metric formalism, this model has been exhaustively explored in inflationary cosmological scenarios \cite{Starobinsky:1982ee,Lyth:1998xn,Carroll:2003wy,DeFelice:2010aj,Clifton:2011jh}, while the Palatini version is known to yield interesting phenomenology involving nonsingular bouncing cosmologies \cite{Barragan:2009sq,Barragan:2010qb}, nonsingular black holes \cite{Olmo:2015axa}, wormholes \cite{Lobo:2020vqh,Bejarano:2017fgz}, and other exotic compact objects \cite{Afonso:2019fzv}. 
When inserted in (\ref{eq:Trace}), this quadratic function leads to the relation $\mathcal{R}=-\kappa T$, exactly like in GR.  
We will refer to the representation (\ref{eq:Ac1}) of the theory as the $f(\mathcal{R})$ frame. Note that in this frame the scalar $\Phi$ is minimally coupled to the metric $g_{\mu\nu}$.

As it was shown in \cite{Afonso:2018hyj}, there exists a correspondence between the theory (\ref{eq:Ac1}) and the Einstein-Hilbert action of the metric $q_{\mu\nu}$ minimally coupled to a matter Lagrangian $K(Z,\Phi)$ (from now on the Einstein frame), namely,  
\begin{equation}
    S_{\rm EH}=\int d^4x \sqrt{-q} \frac{R}{2 \kappa} -\frac{1}{2}\int d^4x \sqrt{-q} K(Z,\Phi) \quad ,
\end{equation}
where the kinetic term $Z=q^{\alpha \beta}\partial_{\alpha}\Phi^*\partial_{\beta}\Phi$ is now contracted with the (inverse) metric $q^{\alpha \beta}$, $R$ is the Ricci scalar of the metric  $q_{\alpha \beta}$, i.e., $R= q^{\alpha \beta}R_{\alpha \beta}(q)$, and $q=\det(q_{\alpha\beta})$. 

For the specified $f(\mathcal{R})$ and $P(X,\Phi)$ functions it can be shown that \cite{Afonso:2018hyj}
\begin{equation}\label{eq:K}
    K(Z,\Phi)=\frac{Z-\xi \kappa Z^2}{1-8 \xi \kappa V}-\frac{2V}{1-8 \xi \kappa V} \quad .
\end{equation}
As we can see, non-linearities in the gravitational sector of the $f(\mathcal{R})$ frame have been transferred into non-linearities in the matter sector of the Einstein frame. Because of this relation between frames, in order to solve the field equations of $f(\mathcal{R})$ gravity coupled to a scalar field  we will solve instead the corresponding problem in GR coupled to the non-linear scalar field matter Lagrangian (\ref{eq:K}). Once the metric $q_{\mu\nu}$ and the scalar field $\Phi$ have been found, we automatically have the metric $g_{\mu\nu}$ via the conformal relation (\ref{eq:conformal}).

\subsection{Evolution formalism}

In order to study the time evolution of boson stars we use the 3+1 Baumgarte-Shapiro-Shibata-Nakamura (BSSN) formalism of Einstein's equations \cite{Baumgarte:1998te, Shibata:1995we} in the Einstein frame. In this formalism space-time is foliated by a family of spatial hypersufaces $\Sigma_t$ labeled by its time coordinate $t$. We denote the (future-oriented) unit normal timelike vector of each hypersurface by  $n^{\alpha}=(1/\alpha,-\beta^i/\alpha)$, and its dual by $n_{\alpha}=(-\alpha,0,0,0)$. Since the system we study has spherical symmetry, the metric in the Einstein frame reads
\begin{equation}\label{eq:EinsteinMetric}
    \begin{aligned}
        ds_{\text{EF}}^2=&-(\alpha^2 -\beta^x \beta_{x})dt^2+2\beta_{x}dx dt \\
        &+ e^{4 \chi}\left( a(t,x)dx^2+x^2 b(t,x) d\Omega^2\right)\quad,\\
    \end{aligned}
\end{equation}
where $d\Omega^{2}=d\theta^2+\sin^2\theta d\varphi^2$, $\alpha$ is the lapse function, $\beta^{x}$ the shift vector, $a(t,x)$ and $b(t,x)$ are the metric functions  and $\chi$ is the conformal factor defined by
\begin{equation}
    \chi=\frac{1}{12}\ln(\gamma/\hat{\gamma}) \quad.
\end{equation}
Note that we use $x$ to denote the radial coordinate. In the last equation, $\gamma$ is the determinant of the spacelike induced metric on every hypersuface $\Sigma_t$,
\begin{equation}
    \gamma_{\alpha \beta}=q_{\alpha \beta}+n_{\alpha}n_{\beta} \quad ,
\end{equation}
and $\hat{\gamma}$ is the determinant of the conformal metric. The latter relates to the full 3-metric by
\begin{equation}
    \hat{\gamma}_{ij}=e^{-4 \chi}\gamma_{ij}\quad.
\end{equation}

Initially, the determinant of the conformal metric fulfills the condition that it equals the determinant of the flat metric in spherical coordinates $\hat{\gamma}(t=0)=x^{4}\sin^2\theta$. Moreover, we follow the so called ``Lagrangian'' condition $\partial_{t}\hat{\gamma}=0$.

In the BSSN formalism the evolved fields are the conformally related 3-dimensional metric $a$ and $b$,  the conformal exponent $\chi$, the trace of the extrinsic curvature $K$, the independent component of the traceless part of the conformal extrinsic curvature, $A_a\equiv A^x_{\, x}$, $A_b\equiv A^\theta_{\, \theta}=A^\varphi_{\, \varphi}$ and the radial component of the conformal connection functions $\hat{\Delta}^x\equiv\hat{\gamma}^{mn}(\hat{\Gamma}^x_{mn}-\hat{\Gamma}^x_{mn}(t=0))$ \cite{Escorihuela-Tomas:2017uac,Montero:2012yr}. Explicitly, the BSSN evolution system reads
\begin{equation}
    \begin{aligned}
        \partial_t a= \beta^x\partial_x a + 2a \partial_x\beta^x-\frac{2}{3} a \hat{\nabla}_x \beta^x-2\alpha a A_a \quad,
    \end{aligned}
\end{equation}
\begin{equation}
    \begin{aligned}
        \partial_t b= \beta^x\partial_x b + 2b \frac{\beta^x}{x}-\frac{2}{3} b \hat{\nabla}_x \beta^x-2\alpha b A_b \quad,
    \end{aligned}
\end{equation}
\begin{equation}
    \begin{aligned}
        \partial_t \chi= \beta^x\partial_x \chi +\frac{1}{6} \left(\alpha K-\hat{\nabla}_x \beta^x\right) \quad ,
    \end{aligned}
\end{equation}
\begin{equation}
    \begin{aligned}
        \partial_t K=& \beta^x\partial_x K - \nabla^2 \alpha+ \alpha (A_a^2+2A_b^2 +\frac{1}{3}K^2) \\
        & +4\pi \alpha \left( \rho +S_a+2S_b\right)\quad ,
    \end{aligned}
\end{equation}
\begin{equation}
    \begin{aligned}
        \partial_t A_a=& \beta^x\partial_x A_a -\left(\nabla^x\nabla_x\alpha-\frac{1}{3}\nabla^2 \alpha\right)+\alpha \left(R^x_x-\frac{1}{3}R\right)\\
        &+a K A_a-16 \pi \alpha (S_a-S_b)\quad,
    \end{aligned}
\end{equation}
\begin{equation}
    \begin{aligned}
        \partial_t \hat{\Delta}^x=& \beta^x\partial_x \hat{\Delta}^x -\hat{\Delta}^x\partial_x\beta^x+\frac{1}{a}\partial^2_x\beta^x +\frac{2}{b}\partial_x\left(\frac{\beta^x}{x}\right)\\
        &+\frac{1}{3}\left(\frac{1}{a}\partial_x(\hat{\nabla}_x \beta^x)+2\hat{\Delta}^x\hat{\nabla}_x\beta^x\right)\\
        &-\frac{2}{a}\left(A_a\partial_x\alpha+\alpha \partial_x A_z\right)\\
        &+2 \alpha \left(A_a \hat{\Delta}^x-\frac{2}{x b}(A_a-A_b)\right)\\
        &+\frac{\xi \alpha}{a}\left[\partial_x A_a-\frac{2}{3}\partial_xK+6 A_a \partial_x \chi\right.\\
        &\left.+(A_a-A_b)\left(\frac{2}{x}+\frac{\partial_x b}{b}\right)-8 \pi j_x\right] \quad .
    \end{aligned}
\end{equation}

When performing the time evolution of the above functions we have to specify a stress-energy tensor and its 3+1 projection. The case we are concerned with is a boson star in Palatini $f(\mathcal{R})=\mathcal{R}+\xi\mathcal{R}^2$ gravity. We write its  stress-energy tensor in the Einstein frame as 

\begin{equation}\label{tensorem}
    \begin{aligned}
        \tilde{T}_{\mu \nu}=&-\frac{2}{\sqrt{-q}}\frac{\partial(\sqrt{-q} K(Z,\Phi))}{\partial q^{\mu \nu}}\\
        =&\frac{1}{2(1+4\xi \kappa \mu^2 |\Phi|^2)}\left[\partial_{\mu}\Phi^*\partial_{\nu}\Phi +\partial_{\nu}\Phi^*\partial_{\mu}\Phi \right.\\
        &\left.-q_{\mu \nu}\partial^{\alpha}\Phi^*\partial_{\alpha}\Phi -\mu^2 q_{\mu \nu} |\Phi|^2 \right. \\
        &\left. - 2\xi \kappa  \partial^{\alpha}\Phi^*\partial_{\alpha} \left(\partial_{\mu}\Phi^*\partial_{\nu}\Phi +\partial_{\nu}\Phi^*\partial_{\mu}\Phi \right) \right.\\
        &\left.+\xi \kappa q_{\mu \nu } \partial^{\alpha}\Phi^*\partial_{\alpha}\Phi\partial^{\beta}\Phi^*\partial_{\beta}\Phi\right]\quad.
    \end{aligned}
\end{equation}
The projections are performed using the unit normal vector $n^\alpha$ and the induced metric $\gamma^{\alpha \beta}$. The matter source terms appearing in the BSSN evolution equations are:
\begin{equation}
    \begin{aligned}
        \rho=&n^{\mu}n^{\nu}\tilde{T}_{\mu \nu}\\
        =&\frac{1}{2(1+4\kappa \xi \mu^2 |\Phi|^2 )}\left[|\Pi|^2+\frac{|\Psi|^2}{ae^{4\chi}}+\mu^2|\Phi|^2 \right.\\
        &\left.-\kappa \xi\left(\frac{|\Psi|^2}{ae^{4\chi}}\right)^2+3\kappa\xi|\Pi|^4-2\kappa \xi \frac{|\Psi|^2}{ae^{4\chi}}|\Pi|^2 \right]\quad ,\\ 
    \end{aligned}
\end{equation}

\begin{equation}
    \begin{aligned}
        S_{a}=&\gamma^{x \mu}\tilde{T}_{x \mu}\\
        =&\frac{1}{2(1+4\kappa \xi \mu^2 |\Phi|^2 )}\left[|\Pi|^2+\frac{|\Psi|^2}{ae^{4\chi}}-\mu^2|\Phi|^2 \right.\\
        &\left. -3\kappa \xi\left(\frac{|\Psi|^2}{ae^{4\chi}}\right)^2+\kappa\xi|\Pi|^4+2\kappa \xi \frac{|\Psi|^2}{ae^{4\chi}}|\Pi|^2 \right]\quad,\\
    \end{aligned}
\end{equation}

\begin{equation}
    \begin{aligned}
        S_{b}=&\gamma^{\theta \mu}\tilde{T}_{\theta \mu}\\
        =&\frac{1}{2(1+4\kappa \xi \mu^2 |\Phi|^2 )}\left[|\Pi|^2-\frac{|\Psi|^2}{ae^{4\chi}}-\mu^2|\Phi|^2 \right. \\
        &\left.+\kappa \xi\left(\frac{|\Psi|^2}{ae^{4\chi}}\right)^2+\kappa\xi|\Pi|^4-2\kappa \xi \frac{|\Psi|^2}{ae^{4\chi}}|\Pi|^2 \right]\quad,\\
    \end{aligned}
\end{equation}

\begin{equation}
    \begin{aligned}
        j_{x}=&-\gamma_{x }^{\mu}n^{\nu}\tilde{T}_{\mu \nu}\\
        =&\frac{1}{2(1+4\kappa \xi \mu^2 |\Phi|^2 )}\left[\frac{1}{a e^{4\chi}}\left(\Pi\Psi^*+\Pi^*\Psi\right)\right.\\
        &\left.+\frac{2\kappa \xi |\Psi|^2}{a^2 e^{8\chi}}\left(\Pi\Psi^*+\Pi^*\Psi\right)-\frac{2\kappa \xi |\Pi|^2}{a e^{4\chi}}\left(\Pi\Psi^*+\Pi^*\Psi\right)\right]\quad .
    \end{aligned}
\end{equation}

Correspondingly, the equations of motion for the scalar field are obtained by reformulating the Klein-Gordon equation in terms of the following two first-order variables
\begin{equation}
    \begin{aligned}
        \Psi& := \partial_{x}\Phi\quad ,\\
        \Pi &:= n^{\alpha}\partial_{\alpha}\Phi=\frac{1}{\alpha}\left( \partial_{t}\Phi-\beta^{x}\Psi\right)\quad .
    \end{aligned}
\end{equation}

In this way the equations of motion for the scalar field read	
\begin{equation}
    \partial_{t}\Phi=\beta^{x}\partial_{x}\Phi+\alpha \Pi\quad,
\end{equation}
\begin{equation}
    \partial_{t}\Psi=\beta^{x}\partial_{x}\Psi+\Psi\partial_{x}\beta^x+\partial_{x}\left(\alpha \Pi\right)\quad,
\end{equation}

\begin{equation}\label{dPidt}
    \partial_{t}\Pi=\frac{1-2\kappa \xi Z+\kappa \xi |\Pi|^2}{1-2\kappa \xi Z+2\kappa \xi |\Pi|^2}\left\{\Xi-\frac{\kappa \xi\Pi^2\Xi^*}{1-2\kappa \xi Z+\kappa \xi |\Pi|^2}\right\}\,,
\end{equation}

\begin{widetext}
    where we have introduced the new variable $\Xi$ in order to simplify the notation, defined as
    \begin{equation}
        \begin{aligned}
            \Xi:=&\beta^{x}\partial_{x}\Pi+\frac{\Psi}{ae^{4\chi}}\partial_{x}\alpha+\frac{\alpha}{a e^{4\chi}}\left[\partial_{x}\Psi+\Psi\left(\frac{2}{x}-\frac{\partial_{x}a}{2a}+\frac{\partial_{x}b}{b}+2\partial_{x}\chi\right) \right]+\alpha K \Pi \\
            &\left.-\frac{\alpha \mu^2 \Phi}{1-2\kappa \xi Z}  +\frac{\alpha \left(Z-\kappa \xi Z^2+\mu^2|\Phi|^2\right)4\xi\kappa\Phi\mu^2}{\left(1+4\kappa \xi \mu^2 |\Phi|^2\right)\left(1-2\kappa \xi Z\right)} \right.\\
            &\left.-\frac{4\kappa \xi \mu^2 \alpha}{1+4\kappa \xi \mu^2 |\Phi|^2}\left[-\frac{\Pi}{\alpha}\left(\partial_{t}\Phi^*\Phi+\Phi^*\partial_{t}\Phi\right)+\left(\frac{\Psi}{e^{4\chi}a}+\frac{\Pi\beta^x}{\alpha}\right)\left(\partial_{x}\Phi^*\Phi+\Phi^*\partial_{x}\Phi\right)\right]\right.\\
            &\left.+\frac{\alpha \kappa \xi}{1-2\kappa \xi Z}\left[\frac{\left(\partial_{t}\Psi^*\Psi+\Psi^*\partial_{t}\Psi\right)e^{4\chi}a-|\Psi|^2\left(4e^{4\chi}a\partial_{t}\chi+e^{4\chi}\partial_{t}a\right)}{e^{8\chi}a^2}\frac{\Pi}{\alpha}\right.\right.\\
            &\left.\left.\textcolor{white}{-\frac{\alpha \kappa \xi}{1-2\kappa \xi Z}\left[\right.}+\left(\frac{\Psi}{e^{4\chi}a}+\frac{\Pi\beta^x}{\alpha}\right)\frac{\left(\partial_{x}\Psi^*\Psi+\Psi^*\partial_{x}\Psi\right)e^{4\chi}a-|\Psi|^2\left(4e^{4\chi}a\partial_{x}\chi+e^{4\chi}\partial_{x}a\right)}{e^{8\chi}a^2}\right.\right.\\
            &\left.\left.\textcolor{white}{-\frac{\alpha \kappa \xi}{1-2\kappa \xi Z}\left[\right.}+\left(\frac{\Psi}{e^{4\chi}a}+\frac{\Pi\beta^x}{\alpha}\right)\left(\partial_{x}\Pi^*\Pi+\Pi^*\partial_{x}\Pi\right)\right] \right. \quad,
        \end{aligned}
    \end{equation}
\end{widetext}
and
\begin{equation}
    Z \equiv q^{\mu \nu}\partial_{\mu}\Phi^*\partial_{\nu}\Phi = \frac{|\Psi|^2}{e^{4\chi}a}-|\Pi|^2 \qquad.
\end{equation}

Within the BSSN formalism we have gauge freedom to choose the ``kinematical variables'', i.e.~the lapse function and the shift vector. As customary in Numerical Relativity, we choose the so-called ``non-advective 1+log'' condition for the lapse function \cite{Bona:1997hp}, and a variation of the ``Gamma-driver'' condition for the shift vector \cite{Alcubierre:2002kk,Alcubierre:2011pkc},
\begin{equation}\label{eq:gauge}
	\begin{aligned}
		\partial_t \alpha &=
  -2 \alpha K\quad,\\
		\partial_t B^x&=
  \frac{3}{4}\partial_t\hat{\Delta}^x
  \quad,\\
		\partial_t \beta^x&=
  B^x \,\,.
  	\end{aligned}
\end{equation}

We also provide the explicit form of the conformal factor $f_\R$. From the Einstein field equations of the Palatini quadratic $f(\R)$ model it can be shown that $\R=-\kappa T$. Therefore,
\begin{equation}
    f_{\R}=1+2\xi\kappa\R
    =\frac{1-8 \kappa \xi V}{1-2\kappa \xi Z}\,\,.
\end{equation}

In addition to the evolution equations, the Einstein-Klein-Gordon system also contains the Hamiltonian and momentum constraint equations. 
These equations read
\begin{equation}\label{eq:hamiltonian}
    \mathcal{H}\equiv R-(A_{a}^2+2A_{b}^2)+\frac{2}{3}K^2-2\kappa\rho=0\quad,
\end{equation}
\begin{equation}
    \begin{aligned}
        \mathcal{M}_{x}\equiv& \partial_{x}A_{a}-\frac{2}{3}\partial_{x}K+6A_{a}\partial_{x}\chi\\
        &+(A_{a}-A_{b})\left(\frac{2}{x}+\frac{\partial_{x}b}{b}\right)-\kappa j_{x}=0 \quad.
    \end{aligned}
\end{equation}

The problem in question is set in the $f(\R)$-frame, the time evolution of boson stars in Palatini $f(\R)$ gravity. We use the conformal relation \eqref{eq:conformal} to translate it to the Einstein frame, as can be seen in the energy-momentum tensor modifications  \eqref{tensorem}. In the Einstein frame we are able to use the BSSN formalism to solve the evolution equations and then translate it to the $f(\R)$ frame again. The metric of this frame is
\begin{equation}\label{eq:fRmetric}
    \begin{aligned}
        d\tilde{s}_{f(\mathcal{R})}^2=&-(\tilde{\alpha}^2 -\tilde{\beta}^r \tilde{\beta}_{r})dt^2+2\tilde{\beta}_{r}dr dt \\
        &+ \tilde{A}(t,r)dr^2+\tilde{R}^2(t,r) d\Omega^2\quad ,
    \end{aligned}
\end{equation}
where the radial coordinate is expressed with an $r$ in order to distinguish it from the radial coordinate $x$ of the Einstein frame.

\section{Initial Data}\label{sec:inidata}
In order to compute the time evolution of boson stars within the context of Palatini quadratic $f(\mathcal{R})$ gravity, initial data must be provided. This is achieved by computing static spherically symmetric boson stars, as described in \cite{Maso-Ferrando:2021ngp}. It is noteworthy that in order to use the BSSN formulation, the system must be expressed in the Einstein frame.

The initial data are obtained in polar-areal coordinates, where the line element is given by the expression
\begin{equation}\label{eq:pametric}
    ds_{\text{pa}}^2=-\alpha_{\text{pa}}^2(x')dt^2+\beta_{\text{pa}}^2(x')dx_{\text{pa}}^2+x'^2d\Omega^2 \quad,
\end{equation}
where $\alpha_{\text{pa}}^2$ and $\beta_{\text{pa}}^2$ are the metric functions and should not be confused with the lapse function and shift vector.

\begin{figure}[t!]
    \includegraphics[width=\linewidth]{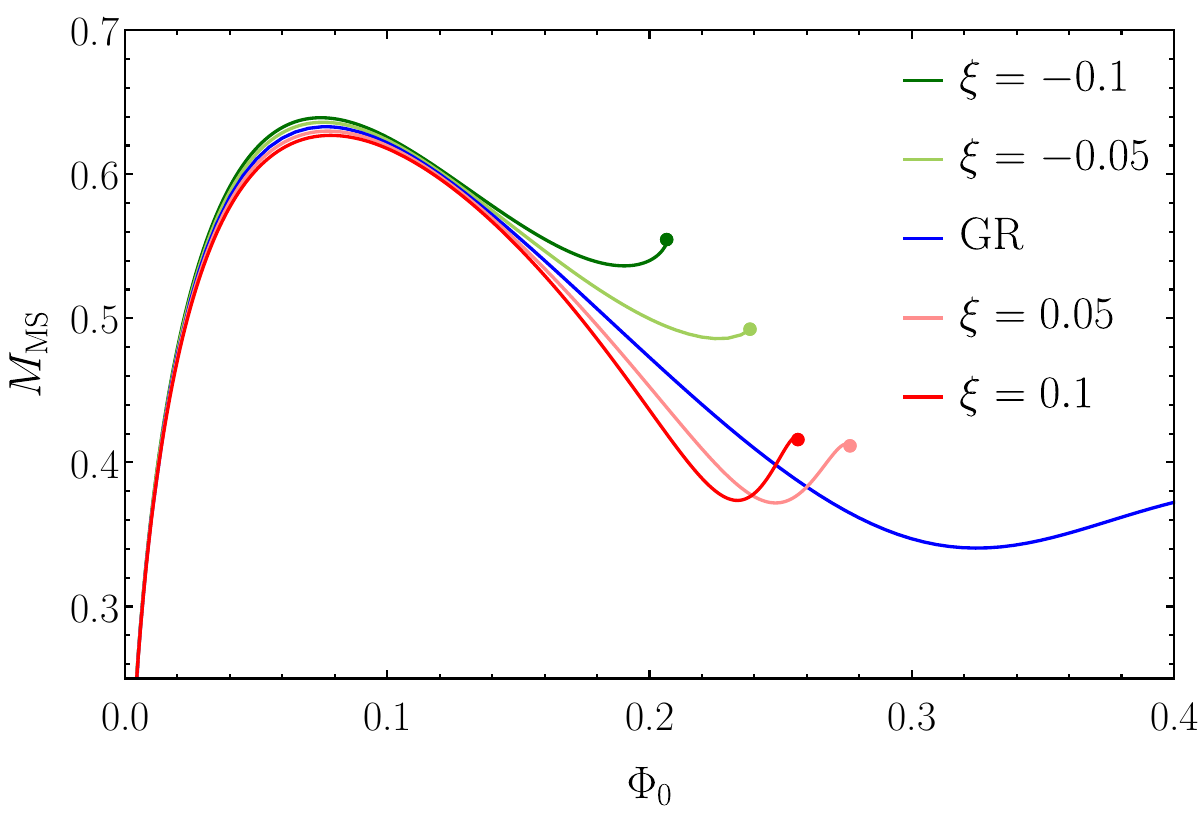}
    \caption{Boson star equilibrium configurations. We represent the mass of different configurations as function of the central value of the scalar field. The various curves correspond to different values of the coupling parameter $\xi$.  A circle indicates the last computable solution. }
    \label{fig:initialdata}
\end{figure}

\begin{table*}[t]
    \centering
    \begin{tabular*}{0.8\textwidth}{c @{\extracolsep{\fill}} ccccc}
        \hline
        \hline
        \text{Model} & $\xi$	& $\Phi_0(t=0)$ & $\omega$ &  $M_{\text{MS}}$   &$E_B=M_{\text{MS}}-\mu N_{EF}$ \\
        \hline
        A(n) & -0.1	& 0.02 & 0.95392  & 0.47925 & -0.00692\\
        A(z) & 0.0	& 0.02 & 0.95419 & 0.47514 &-0.00679\\
        A(p) & 0.1	& 0.02 & 0.95445 &  0.47108 &-0.00665\\
        \hline
        B(n) & -0.1	& 0.1  & 0.82241 & 0.62571 & -0.01758\\
        B(z) & 0.0	& 0.1  & 0.82296 & 0.62180  &-0.01775\\
        B(p) & 0.1	& 0.1  & 0.82350 &  0.61787 &-0.01790\\
        \hline
        C(n) & -0.1	& 0.18 & 0.75311  & 0.53922  &0.00576\\
        C(z) & 0.0	& 0.18 & 0.76904  & 0.50671  &0.01353\\
        C(p) & 0.1	& 0.18 & 0.77840 & 0.48574  &0.01780\\
        \hline
    \end{tabular*}
    \caption{\label{tab:taula1}Parameters for our nine initial boson star configurations. From left to right each column reports the model name, its gravitational coupling factor, the initial value of the central scalar field (i.e.~at $r=x=0$), its frequency, the Misner-Sharp mass associated of the configuration, and the binding energy in both frames. The letters `n', `z', and `p', stand for the negative, zero, and positive values of the coupling parameter, respectively.}
\end{table*}

To solve for the static configurations of boson stars, it is assumed that the scalar field can be expressed as $\Phi(x_{\text{pa}},t)=\phi(x_{\text{pa}})e^{i \omega t}$, where $\phi(x_{\text{pa}})$ is the radial distribution of the scalar field and $\omega$ is the frequency. The Einstein-Klein-Gordon system is then derived. The integration is performed with appropriate boundary conditions, ensuring regularity at the origin and asymptotic flatness. A fourth-order Runge-Kutta scheme with adaptive step size and a shooting method is employed, leaving $\Phi_0\equiv\phi(x_{\text{pa}}=0)$ as a free parameter. The grid used to compute the initial data is an equidistant grid with spatial resolution $\Delta x_{\text{pa}}=0.0025$. By solving this system, the metric functions $\alpha_{\text{pa}}^2$ and $\beta_{\text{pa}}^2$, as well as the frequency $\omega$ and the radial distribution of the scalar field $\phi(x_{\text{pa}})$, can be obtained. This results in a collection of static configurations of boson stars, each described by a different value of $\Phi_0$. They are plotted in Figure \ref{fig:initialdata}. We show mass profiles as function of the central scalar field, $\Phi_0$, for five different values of the gravitational coupling parameter $\xi$, two of them are positive, two negative and the zero value which is equivalent to GR.

The mass of the configurations is computed using the Misner-Sharp expression, a well-established mathematical formula that quantifies the mass from the point of view of a distant observer,
\begin{equation}
    M_{\text{MS}}=\frac{x^{\text{max}}_{\text{pa}}}{2}\left(1-\frac{1}{\beta_{\text{pa}}^2(x^{\text{max}}_{\text{pa}})}\right)\quad.
\end{equation}
Notably, we find that the computed mass remains consistent in both frames. This is because the Misner-Sharp expression captures the mass that a distant observer would perceive, and when observations are made far away from the matter sources, the frames are effectively indistinguishable in terms of the computed mass.

The determination of the number of particles in the system involves two distinct definitions depending on the chosen frame of reference. When computed using the $f(\mathcal{R})$ frame, the number of particles  derived from the conserved quantity that arises from the U(1) symmetry of the scalar field is 
\begin{equation}
    N_{f(\mathcal{R})}=4\pi\int_{0}^{\infty} \frac{dx_{\text{pa}}x_{\text{pa}}^2}{f_\mathcal{R}^{3/2}} \omega \frac{\phi^2 \beta_{\text{pa}}}{\alpha_{\text{pa}}}\left(1-\frac{x_{\text{pa}}}{2 f_\mathcal{R}}\frac{\partial f_{\mathcal{R}}}{\partial x}\right)\quad .
\end{equation}
On the other hand, if the number of particles is computed for the case of GR coupled to a non-linear scalar field matter Lagrangian, the expression for $N_{\text{EF}}$ becomes 
\begin{equation}
    N_{\text{EF}}=4\pi\int_{0}^{\infty} \frac{dx_{\text{pa}}x^2_{\text{pa}}}{f_\mathcal{R}} \omega \frac{\phi^2 \beta_{\text{pa}}}{\alpha_{\text{pa}}}\quad .
\end{equation}

The binding energy $E_B$, which is a crucial parameter in determining the fate of the boson star, can be calculated using the number of particles in the Einstein frame, as this is the frame where the evolution of the system will be performed. Specifically, the binding energy is given by $E_B = M_{\text{MS}} - \mu N_{\text{EF}}$. 

The mass of a boson star is a crucial factor in determining its ultimate fate. Boson star configurations with a central field value $\Phi_0$ lower than $\Phi_0(M^{\text{max}}_{\text{MS}})$  are expected to be stable over time, where $M^{\text{max}}_{\text{MS}}$ represents the maximum mass of the family of boson stars configurations with the same gravitational coupling $\xi$, that is, each of the curves displayed in Figure~\ref{fig:initialdata}. On the other hand, configurations with a central field value $\Phi_0$ higher than $\Phi_0(M^{\text{max}}_{\text{MS}})$ are expected to be unstable. For the latter case, the fate of unstable boson stars depends on its binding energy. Specifically, the binding energy will determine whether the unstable configuration migrates to a stable one ($E_B<0$) or if it disperses ($E_B>0$). The interplay between the maximum mass and binding energy is critical in understanding the long-term stability and dynamical behavior of boson stars.

Nine different configurations are studied in this work, their initial parameters shown in Table \ref{tab:taula1}. Models A are expected to be stable under small linear perturbations while models B and C are unstable. For each set, we will evolve in time three models with the same boson star parameters but in the context of three different gravitational scenarios given by $\xi=\left\{-0.1, 0\equiv \text{GR}, 0.1\right\}$. Our choice of the magnitude for the gravitational coupling parameter, $|\xi|=0.1$, is because such a value is high enough to make visible any differences with respect to GR while being one order of magnitude suppressed. Configurations with any other value for $|\xi|$ would have experienced different behavior quantitatively but not qualitatively.

\section{Numerical Framework}\label{sec:numerical}

The initial boson star configurations are obtained in polar-areal coordinates while the time evolution is carried out in isotropic coordinates using the  numerical-relativity code \texttt{NADA1D}~\cite{Montero:2008yx}. Therefore, a change of coordinates is necessary. By comparing equations \eqref{eq:EinsteinMetric} and \eqref{eq:pametric}, we can deduce that
\begin{equation}
    \beta_{\text{pa}}^2(x_\text{pa}) dx_\text{pa}^2=e^{4\chi(t,x)}a(t,x)dx^2 \qquad ,
\end{equation}
\begin{equation}
    x_\text{pa}^2=e^{4\chi(t,x)}b(t,x)x^2\qquad .
\end{equation}
Here, $x_{\text{pa}}$ and $x$ represent the radial coordinates in polar-areal coordinates and isotropic coordinates, respectively.
Since the change of coordinates is performed before the time evolution begins, i.e., at $t=0$, the metric functions can be set as $a(0,x)=b(0,x)=1$. Combining the two previous equations, we obtain
\begin{equation}\label{canvicoor}
    \frac{dx}{dx_{\text{pa}}}=\beta_{\text{pa}}(x_{\text{pa}})\frac{x}{x_{\text{pa}}} \quad .
\end{equation}
From the fact that the spacetime resembles the Schwarzschild spacetime far away from the object, we can deduce that
\begin{equation}
    x^{\text{max}}=\left[\left(\frac{1+\sqrt{\beta_{\text{pa}}(x_{\text{pa}}^{\text{max}})}}{2}\right)^2 \frac{x_{\text{pa}}^{\text{max}}}{\beta_{\text{pa}}(x_{\text{pa}}^{\text{max}})}\right] \quad ,
\end{equation}
which will be used as the initial value to solve equation \eqref{canvicoor}. For further details about this calculation, we refer the reader to Appendix D of \cite{Lai:2004fw}. Upon establishing the change of coordinates, we can then proceed to calculate the initial conformal factor $e^{4\chi}$ in isotropic coordinates, which is given by the expression
\begin{equation}
    e^{4\chi(0,x)}=\left(\frac{x_{\text{pa}}}{x}\right)^2 \quad.
\end{equation}
This allows us to establish the relationship between the conformal factor and the radial coordinates at the initial state of the system.

Once the coordinate transformation has been carried out, we can determine the initial values of the scalar field quantities in isotropic coordinates. Specifically, we obtain the values of $\Phi(t=0,x)$, $\Psi(t=0,x)$, and $\Pi(t=0,x)$.

After transforming the polar-areal grid into an isotropic grid we interpolate with a cubic-spline over the radial coordinate in order to have the initial configuration on a grid composed of two patches. This grid consists on a geometrical progression in the interior part up to a given radius and a hyperbolic cosine outside. Details about the computational grid can be found in \cite{Sanchis-Gual:2015sxa}. For the logarithmic grid the minimum resolution used is $\Delta x=0.025$. With this choice the inner boundary is then set to $x_{\text{min}}=0.0125$ and the outer boundary is placed at $x_{\text{max}}=8000$. The time step is given by $\Delta t =0.3\Delta x$ in order to obtain long-term stable simulations.

The BSSN equations are solved numerically using a second-order Partially Implicit Runge-Kutta scheme \cite{Cordero2012,Cordero2014}, as implemented in the \texttt{NADA1D} code~\cite{Montero:2008yx}. This scheme can handle in a satisfactory way the singular terms that appear in the evolution equations due to our choice of curvilinear coordinates. Further details about the numerical method can be found in \cite{Escorihuela-Tomas:2017uac}.

\section{Results}\label{sec:results}
\subsection{Stable models}

\begin{figure}[h]
    \centering
    \includegraphics[width=\linewidth]{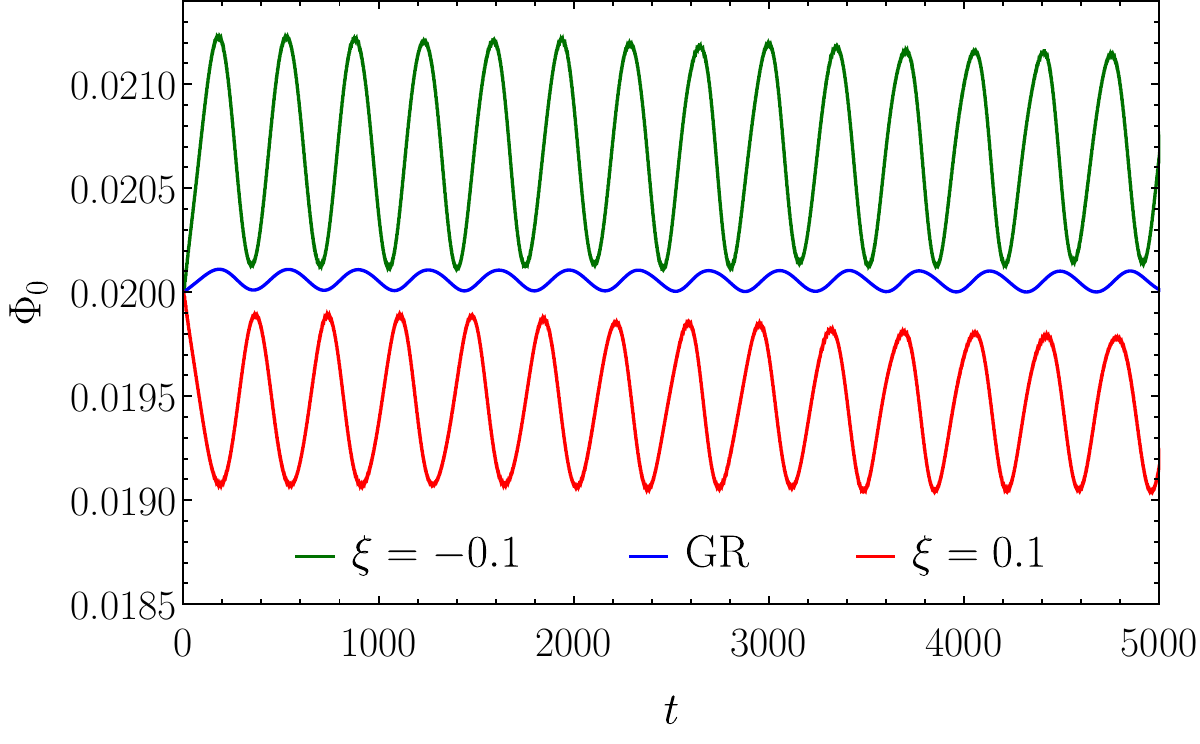}
    \caption{Time evolution of the central value of the scalar field for the models A(n) (green curve), A(z) (blue curve), and A(p) (red curve).}
    \label{fig:modelA}
\end{figure}

The fate of a boson star is determined by the maximum mass of its static configurations in GR, as previously discussed. We find that in Palatini $f(\R)$ theory the same criterion holds. More precisely, initial configurations with a central value of the scalar field lower than $\Phi_0(M^{\text{max}}_{\text{MS}})$ are expected to exhibit stable evolution. 

\begin{figure}[t]
    \centering
    \includegraphics[width=\linewidth]{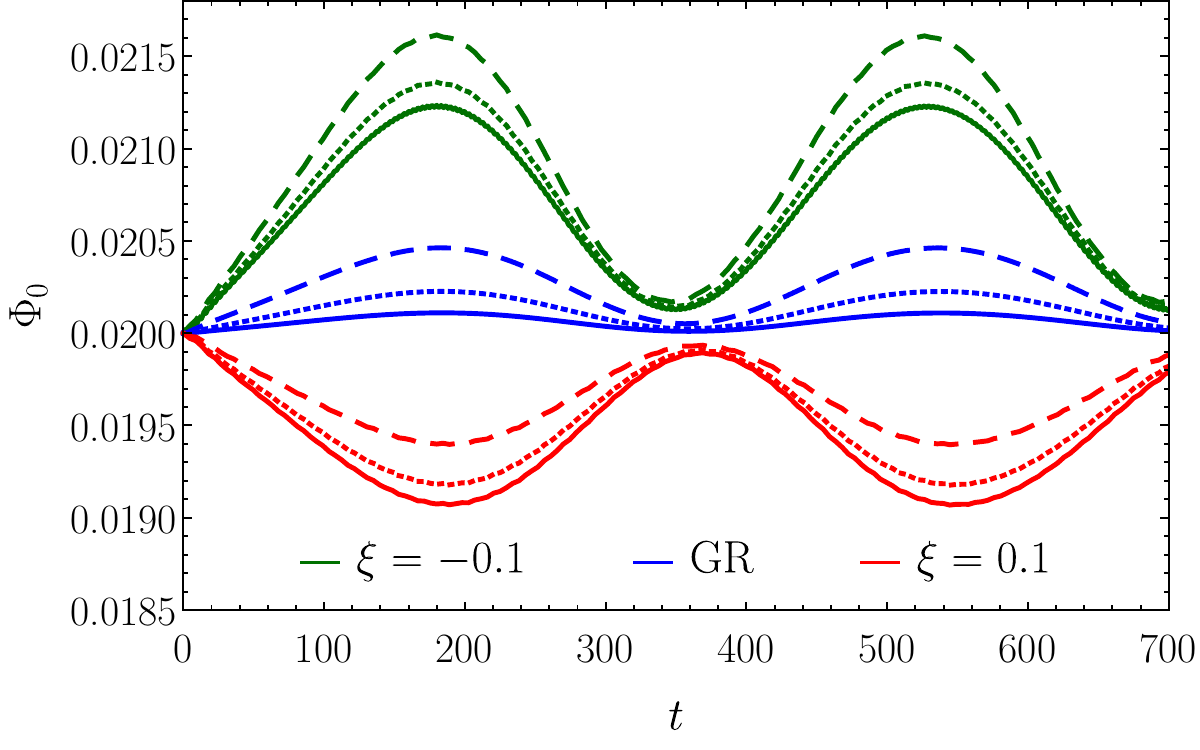}
    \caption{Comparison of the time evolution of the scalar field central value for models A(n) (green lines), A(z) (blue lines) and A(p) (red lines) with three different grid resolutions for the initial data. Solid lines correspond to $\Delta x_{\text{pa}}=0.0025$, dotted lines to $\Delta x_{\text{pa}}=0.005$ and dashed lines to $\Delta x_{\text{pa}}=0.01$.}
    \label{fig:Aresolutions}
\end{figure}

The time evolution results for models A(n), A(z), and A(p) are depicted in Figure \ref{fig:modelA}. The plot illustrates the temporal behavior of the central value of the scalar field, denoted as $\Phi_0(t)\equiv \sqrt{\operatorname{Re}[\Phi(x=0,t)]^2+\operatorname{Im}[\Phi(x=0,t)]^2}$. Notably, considering that $f_{\R}(x=0)\neq0$ and based on the conformal relation between metrics given by Eq.~\eqref{eq:conformal}, it follows that $\Phi_0\equiv\Phi(x=0)=\Phi(r=0)$. Despite all three configurations having the same initial value for the scalar field at the center, namely $\Phi_0(t=0)=0.02$, the frequencies of the scalar field differ due to  the distinct gravitational theories in which they are described, as shown in Table \ref{tab:taula1}. The discrepancies are notably larger for models C.

In the context of GR, i.e.~in the evolution of the model A(z), it is expected to observe a stable boson star, with the central value of the scalar field remaining constant (see e.g.~\cite{Guzman:2009xre,Escorihuela-Tomas:2017uac}). However, due to discretization errors associated with the numerical grid used in the time evolution, all physical quantities, including the central value of the scalar field $\Phi_0$, exhibit instead small-amplitude oscillations around an equilibrium value. With the particular resolution used in our simulation the amplitude of these oscillations is found to be $\Delta \Phi=5\times10^{-5}$.

Qualitatively, the same kind of oscillatory behaviour is found in $f(\R)$ gravity. However, interestingly, the amplitudes of the oscillations are significantly larger in those cases (see green and red curves in Fig.~\ref{fig:modelA}). 
For the models A(n) and A(p), the amplitudes are measured to be $\Delta \Phi=6.2\times 10^{-4}$ and $\Delta \Phi=4.7\times 10^{-4}$, respectively. Notably, the amplitude of the oscillations is found to be proportional to the gravitational coupling parameter $\xi$, indicating a dependence on the specific gravity model being considered. Furthermore, there is a phase shift observed in the A(p) model compared to the other two models, causing the oscillations to shift downwards.

\begin{figure}[t!]
    \centering
    \includegraphics[width=\linewidth]{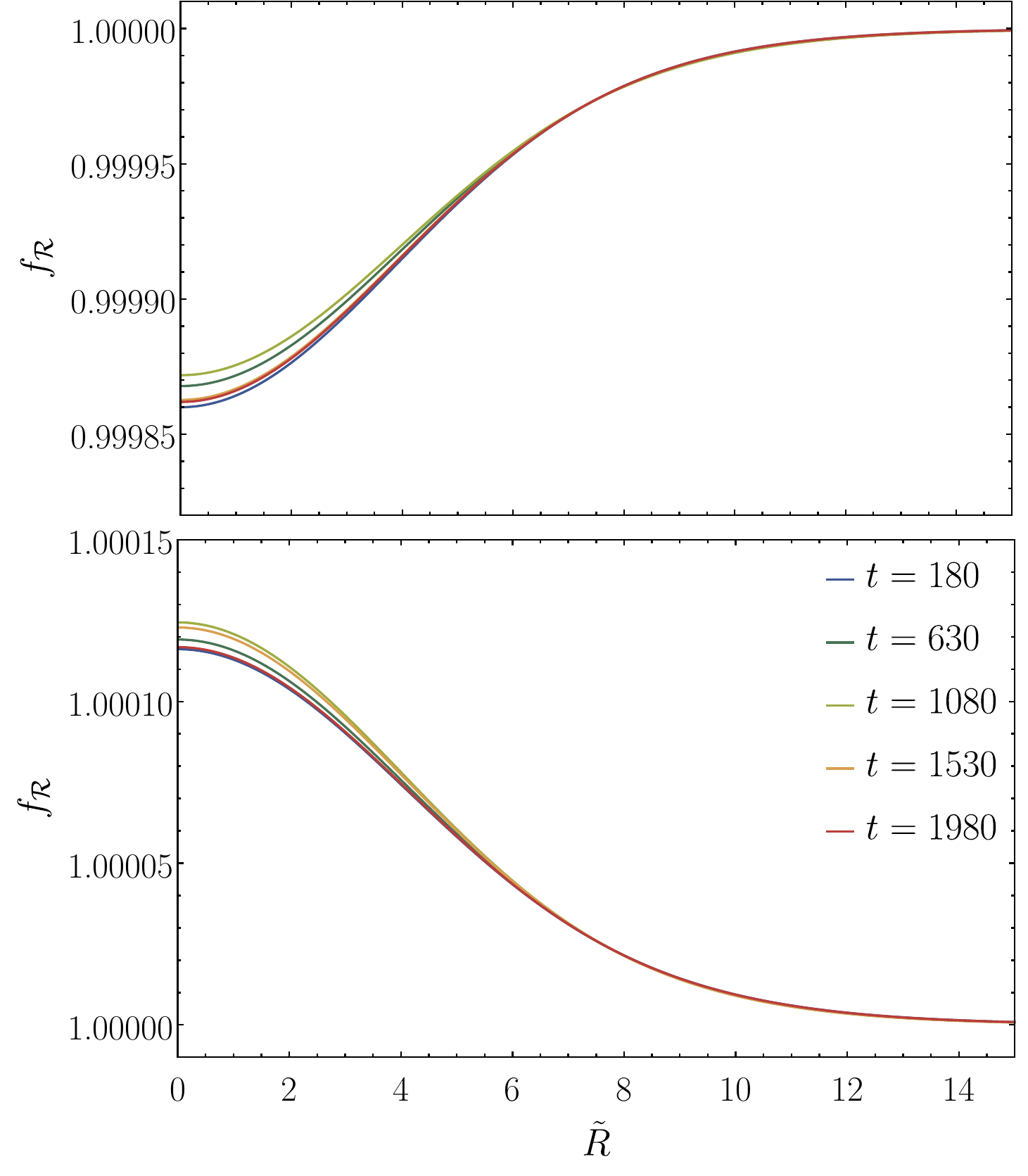}
    \caption{Radial profiles of the conformal factor for models A(n) (upper panel) and A(p) (bottom panel) at selected evolution times.}
    \label{fig:fRRstable}
\end{figure}

\begin{figure}[b!]
    \centering
    \includegraphics[width=\linewidth]{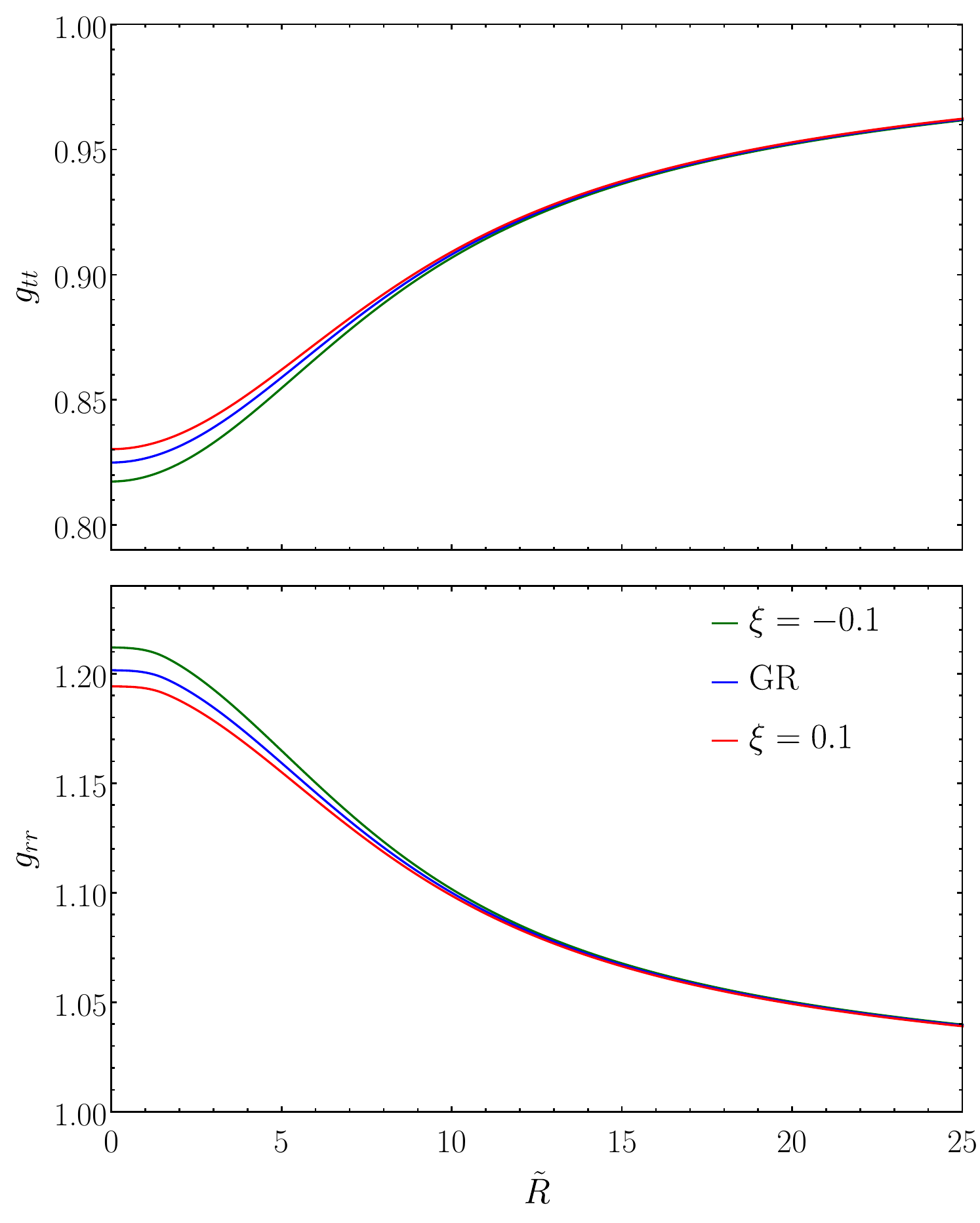}
    \caption{Radial profile of the $g_{tt}$ and $g_{rr}$ metric functions for the models A(n) and A(p) at $t=1575$.}
    \label{fig:metricstable}
\end{figure}

To study the impact of the polar-areal grid resolution on the amplitude of the oscillations, we also performed numerical simulations by systematically varying the resolution of the grid used for computing the initial data. The results are displayed in Figure~\ref{fig:Aresolutions}, which is similar to Figure~\ref{fig:modelA}, but shows data for three different grid resolutions and in a shorter time span. We observe that the amplitude of the oscillations strongly depends on the resolution. From our convergence analysis, for models A(n) and A(p) the oscillation seems to tend to a finite value as the resolution becomes finer rather than disappearing. This is in contrast to GR models, for which the oscillation decreases with resolution as expected. The reason behind this effect is that, when non-linear terms in the matter Lagrangian are present the change of coordinates and subsequent interpolation introduce a larger source of numerical error that we cannot get rid of at these resolutions, which contributes to the amplitude of the mentioned oscillations. However, the qualitative output of the simulation remains unaffected, as the amplitude of the oscillations is only up to 3\% of the total scalar field amplitude for a polar-areal grid resolution $\Delta x_{\text{pa}}=0.0025$. 

By performing several evolutions with different resolutions, we are able to infer the convergence order of the code with respect to the polar areal grid, which is of first order. This loss of convergence is due to the change of coordinates from polar-areal to isotropic, also observed in~\cite{Escorihuela-Tomas:2017uac} (see also the related discussion in~\cite{Maso-Ferrando:2023nju}). Moreover, since we do not further change $\Delta x_{\text{pa}}$ in the simulations, increasing the isotropic grid resolution for the computation of the initial data does not lead to an improved convergence. We refer the reader to Appendix \ref{sec:convergence} for details on the convergence analysis of the evolution code.

Regarding the behaviour of the space-time variables in different theories, we depict in Figure \ref{fig:fRRstable} radial profiles of the conformal factor $f_\R$ for models A(n) and A(p) at selected evolution times. To express the radial position in terms of variables within the $f(\R)$ frame, we employ the area of the two-spheres $\tilde{R}^2$ as a pseudo-coordinate due to the absence of an explicit expression for $r$. As one can observe deviations from unity are only noticeable for points close to the boson star center, where the maximum of the energy density is located, and even in this case it is a minute difference. This suggests that the disparity between the metrics of both frames will be minimal. It can also be noticed that the conformal factor exhibits oscillations of a similar nature as those previously discussed for the maximum of the scalar field $\Phi_0$. The amplitude of these oscillations is about $10^{-5}$. Furthermore, the opposite signs of the coupling parameter $\xi$ affect the radial profile of $f_\R$ in opposite ways for both models. Specifically, the negative sign of $\xi$ (top panel in the figure) tends to enlarge the conformal factor close to the boson star's center, while the opposite effect is observed for $\xi=0.1$.	

\begin{figure*}[t!]
    \includegraphics[width=\linewidth]{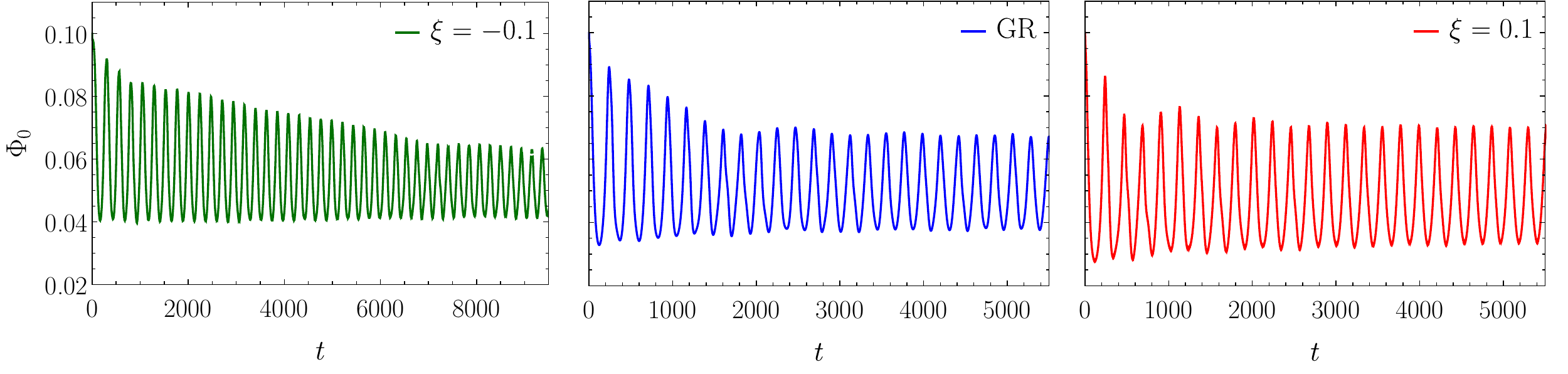}
    \caption{Time evolution of the central scalar-field amplitude for models B(n) (left), B(z) (middle), and B(p) (right). All models experience a migration to the corresponding stable-branch model.}
    \label{fig:modelB}
\end{figure*}

Next, in Figure~\ref{fig:metricstable} we show the radial profiles of the metric functions at $t=1575$. For both models, the metric function $g_{tt}$ starts from a finite positive value below 1, gradually increasing with radial distance and asymptotically approaching 1. As for the function $g_{rr}$, a similar behavior is observed, but with an initial value at the center of the star that is finite and greater than 1 and tending asymptotically toward 1. The discrepancy between the two models becomes visible only close  to the center of the boson star. Though not shown here, these two functions are subject to the aforementioned small oscillations as well.

\subsection{Unstable models}

\begin{figure}[b]
    \centering
    \includegraphics[width=\linewidth]{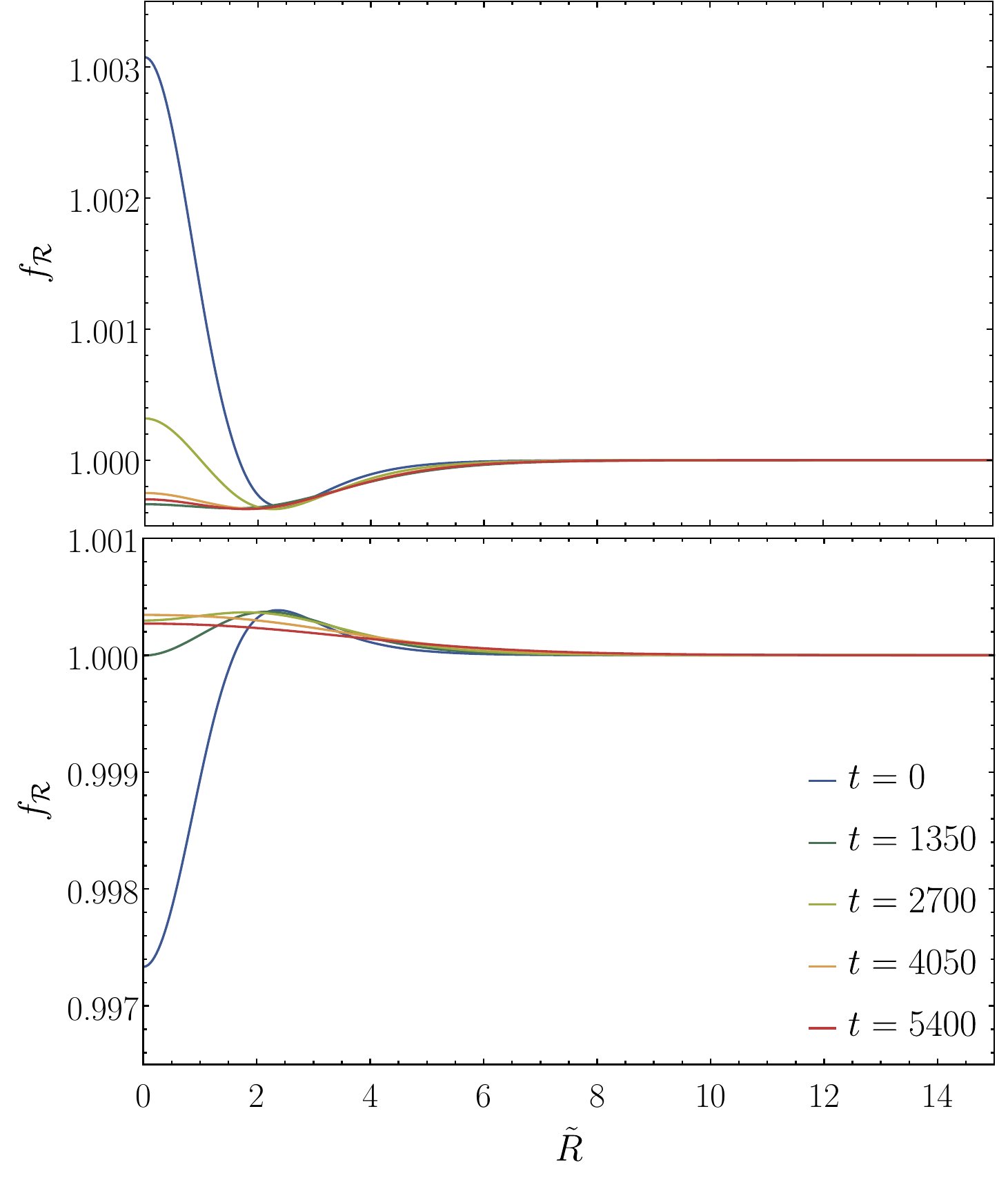}
    \caption{Evolution of the radial profiles of the conformal factor for models B(n) (upper panel) and B(p) (bottom panel).}
    \label{fig:fRRmigration}
\end{figure}

Let us now discuss the temporal evolution of the B(n), B(z), and B(p) models, which are located in the unstable branch and still exhibit a negative binding energy. When the only perturbation to the initial data is the discretization error, we observe a migration of these unstable configurations towards the corresponding boson star with the same mass but located in the stable branch. This behaviour is depicted in Figure \ref{fig:modelB}. The initial central value of the scalar field for all three models is $\Phi_0=0.1$, and it evolves over time until reaching a configuration with $\Phi_0\approx0.055$. As can be inferred from Figure~\ref{fig:initialdata} this value corresponds to stars with approximately the same mass but situated in the stable branch.

\begin{figure}[t]
    \centering
    \includegraphics[width=\linewidth]{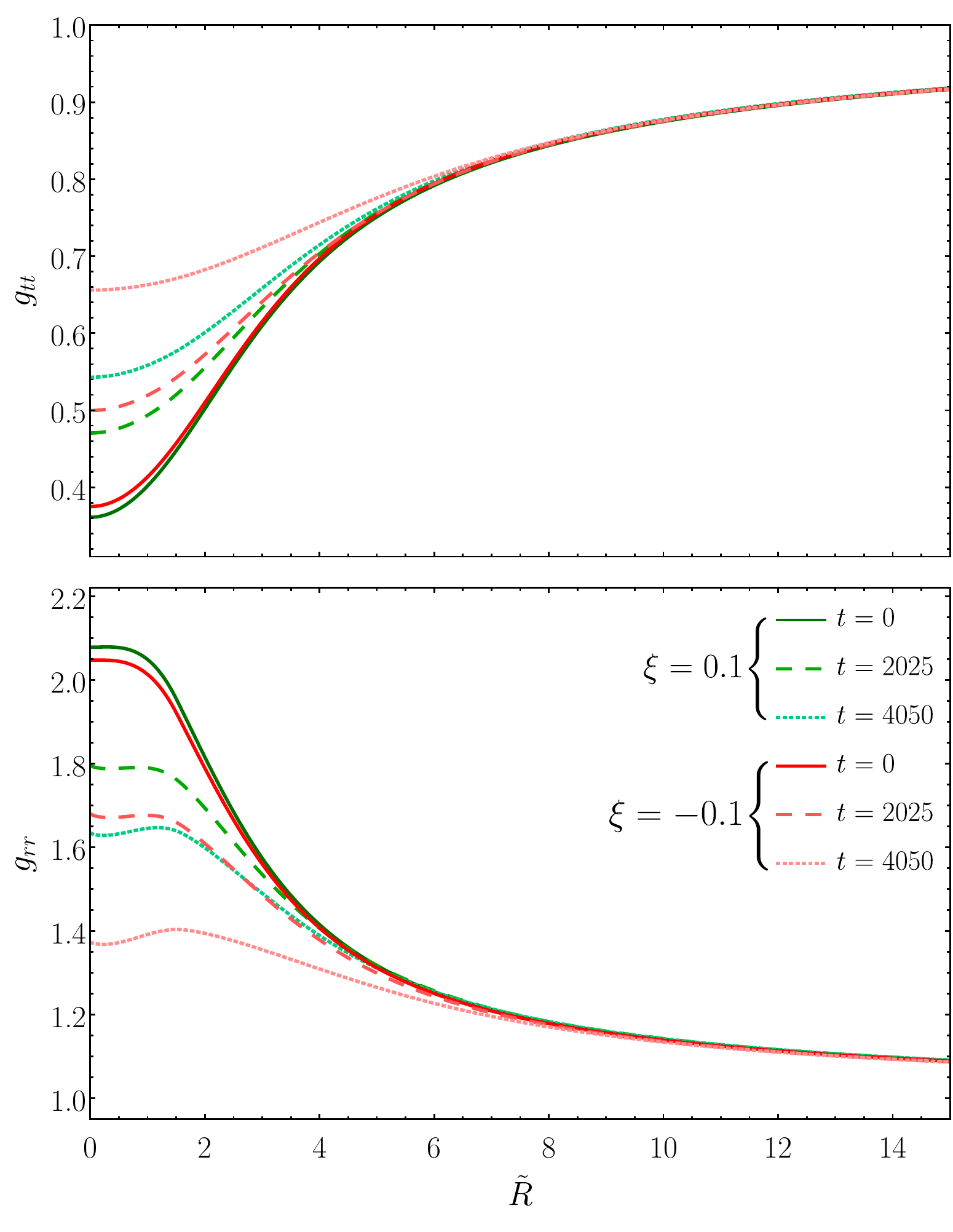}
    \caption{Radial profile of the $g_{tt}$ and $g_{rr}$ metric functions for the models B(n) and B(p).}
    \label{fig:metricmigration}
\end{figure}

In Figure \ref{fig:fRRmigration} we plot radial profiles of the conformal factor $f_\R$ at both the initial time and selected times during the evolution. This figure shows that the initial configuration of the conformal factor exhibits a significant deviation from unity, which gradually diminishes over time. Specifically, for model B(n) (top panel), the value of the conformal factor at the center of the boson star initially exceeds unity but decreases below 1 as the system approaches a stable configuration. Conversely, in the case of model B(p) (bottom panel) the conformal factor follows the opposite trend. However, it is important to note that the conformal factor consistently approaches one asymptotically, either increasing for the B(n) model or decreasing for the B(p) model. Additionally, we show in Figure \ref{fig:metricmigration} the radial profiles of metric functions $g_{tt}$ and $g_{rr}$.  The central values of both metric functions  transition towards one during the evolution. We also note that both the conformal factor and the metric functions exhibit oscillations, which become more apparent when observing the central values over time, as shown in Figure \ref{fig:modelB}.

\begin{figure}[t]
    \centering
    \includegraphics[width=\linewidth]{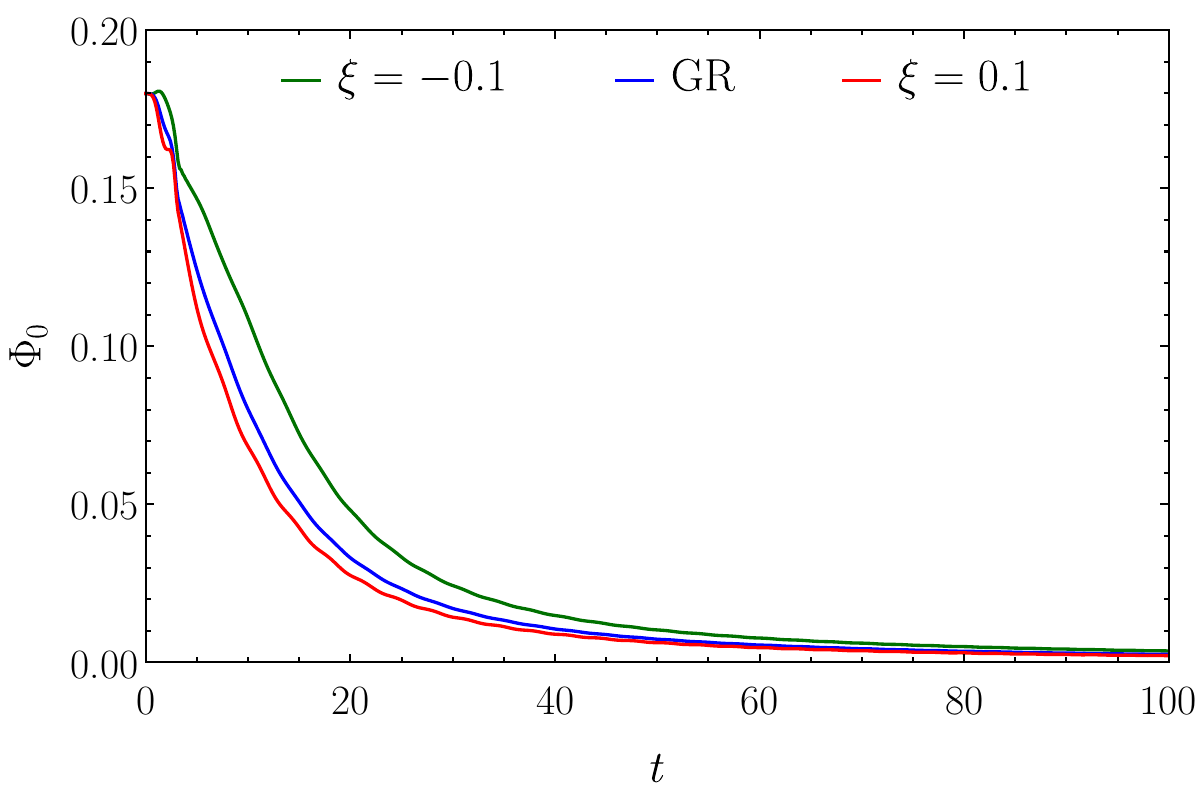}
    \caption{Time evolution of the scalar-field central value for models C(n), C(z), and C(p). All boson stars suffer a total dispersion due to the positive binding energy of the initial data.}
    \label{fig:modelC}
\end{figure}

Turning our attention towards the time evolution of C models, characterized by initial data $\Phi_0>\Phi_0(M^{\text{max}}_{\text{MS}})$ and a positive binding energy $E_B>0$. These models, denoted as C(n), C(z), and C(p), respectively, are representative of different gravitational theories and are also summarized in Table \ref{tab:taula1}. The evolution of the central value of the scalar field, $\Phi_0$, is depicted in Figure \ref{fig:modelC}. It is observed that $\Phi_0$ rapidly decreases with time, leading to a drastic radial expansion of the boson star, which ultimately disperses away. Similar behavior is observed for all three models, although slight quantitative differences exist in the evolution of the central value of the scalar field.

Let us now come back on the B models. However, if we do not rely on discretization error but truly perturb the initial data for the B(n), B(z), and B(p) models, the resulting dynamics can be markedly different. In particular, we can trigger the gravitational collapse of the boson stars, as first shown in~\cite{Maso-Ferrando:2023nju}. To do so, once we have solved the Einstein-Klein-Gordon system, which provides the initial data for the evolution, we multiply the radial profile of the scalar field by 1.02, i.e., we add a 2\% perturbation to this profile. 
This results in a slight violation of the constraints in polar-areal coordinates. After adding the perturbation we do not recompute the spacetime variables $\alpha_\text{pa}$ and $\beta_\text{pa}$. This decision is based on the observation that it only leads to a 3\% increase in the magnitude of the Hamiltonian constraint violation in regions near the center, when compared to the unperturbed case. We note that the introduced perturbation is larger than the one associated with the discretization error, but small enough not to substantially alter our original solution. Once the perturbed scalar field has been obtained, we re-compute the remaining scalar field quantities for the BSSN evolution.

    \begin{figure}[b]
    \centering
    \includegraphics[width=\linewidth]{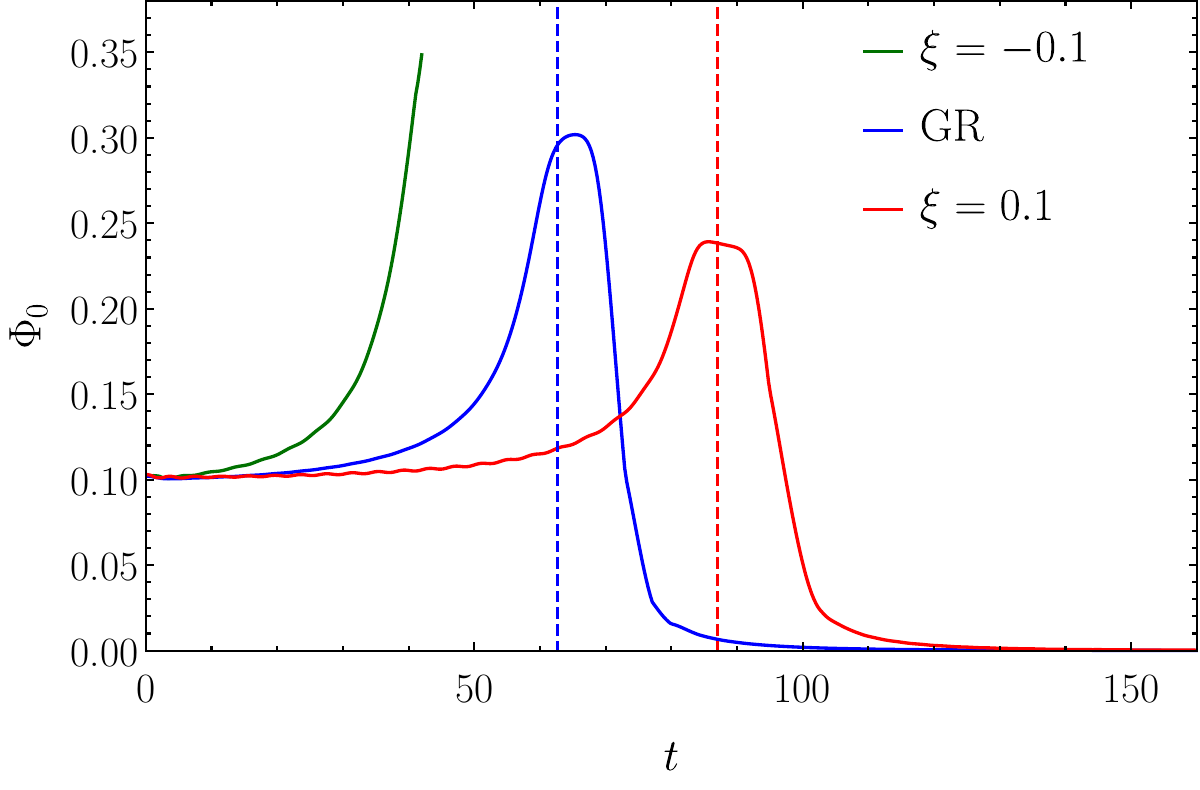}
    \caption{Time evolution of the scalar-field central value for models B(n), B(z), and B(p) after they have been subjected to a 2\% perturbation. The dashed vertical lines indicate the moment in which an apparent horizon forms for each model.}
    \label{fig:modelB_per}
\end{figure}

Figure \ref{fig:modelB_per} shows the evolution of the central value of the scalar field for all three perturbed B models. When evolving these configurations with a perturbation in GR (blue curve), the outcome is the gravitational collapse of the boson star and the formation of a black hole~\cite{Escorihuela-Tomas:2017uac}. The central scalar field is seen to grow up to a maximum value to then decay when an apparent horizon (AH) appears. The AH, signaled with a vertical blue dashed line in Fig.~\ref{fig:modelB_per}, is computed using the AH finder described in~\cite{Thornburg:2006zb}. The mass of the resulting black hole is slightly smaller than the mass of the initial boson star, since some amount of the scalar field is not swallowed by the black hole. This results in a long-lived  cloud of scalar field around the black hole (see~\cite{Escorihuela-Tomas:2017uac} for further details).

Upon analyzing the gravitational collapse of the B(n) model, we observe that after reaching $t=42$, the code stops. If we examine the conformal factor during this evolution, we find that shortly before the code stops it grows rapidly and eventually leads to a divergence. This is due to the fact that the condition $1-2 \kappa \xi Z=0$ is met. Similarly, we find that the equations governing the scalar field evolution also diverge. If we examine Eq.~\eqref{dPidt}, we can see that the combination $1-2\kappa \xi Z$ appears as a denominator. The divergence of $\Pi$ would also induce divergences in $\Phi$ and $\Psi$. Therefore, we are unable to accurately predict the outcome of the gravitational collapse for $\xi=-0.1$ using the formalism presented in this work.

\begin{figure}[t!]
    \centering
    \includegraphics[width=\linewidth]{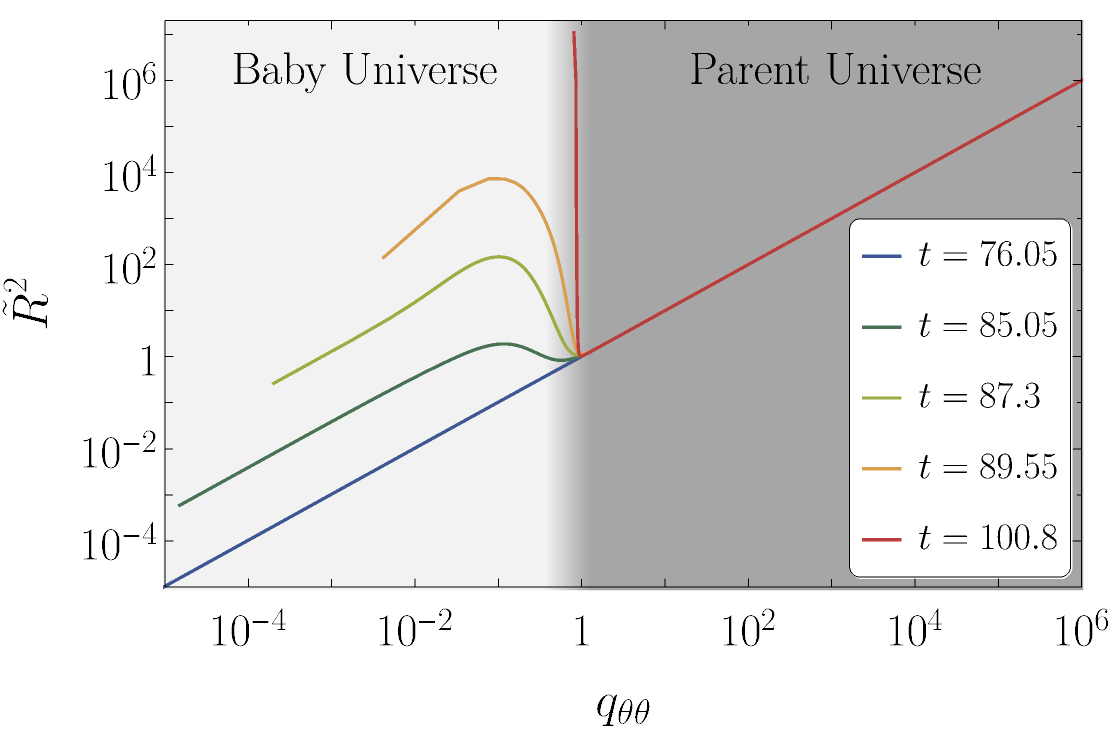}
    \caption{Relationship between the area of the two-spheres in both frames at five selected times. The background indicates the regions referred to as \textit{baby universe} and \textit{parent universe}.}
    \label{fig:Plot2}
\end{figure}

\begin{figure}[b]
    \centering
    \includegraphics[width=\linewidth]{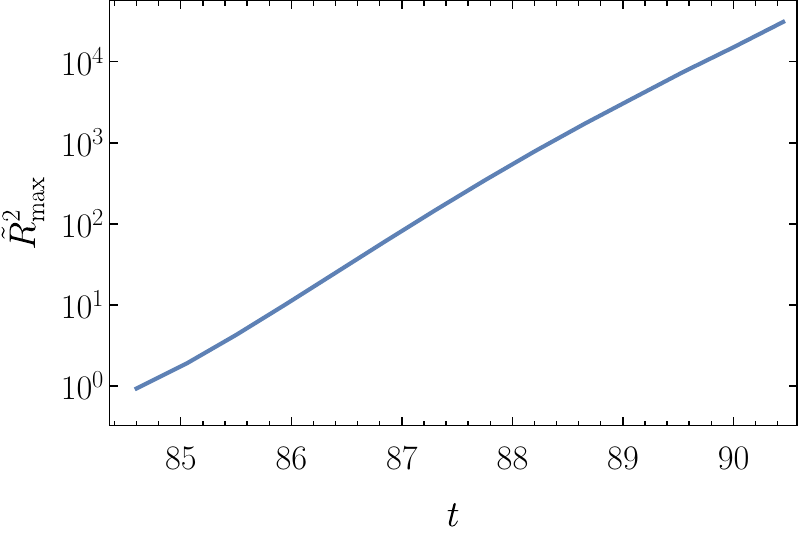}
    \caption{Maximum value of the 2-spheres area $\tilde{R}^2_\text{max}$ for the baby for the expanding baby universe.}
    \label{fig:sizeuni}
\end{figure}

For the B(p) model, the gravitational collapse results in the formation of a black hole surrounded by a cloud of scalar field, similar to the B(z) case (i.e.~GR). However, a glance at the spherical sector of the metric yields crucial new information that highlights this branch of solutions over the others. In fact, it turns out that the relation between the area of the $2$-spheres in the GR and $f(R)$ frames becomes non-monotonical when the collapse sufficiently increases the energy density around the center of the object. This means that as the area of the $2$-spheres in the GR frame decreases as one approaches the center, in the  $f(R)$ frame one observes a transition triggered by the increase in the energy density in which the innermost $2$-spheres experience an inflationary expansion (see Figure~\ref{fig:Plot2}). This is a manifestation of repulsive gravity effects that arise due to the modified gravitational dynamics. When the energy density is sufficiently high, the collapsing field bounces off but since the causal structure prevents the dissipation of the object, the only natural way out is the transition from a collapsing scenario to an expanding one, in much the same way as one finds in nonsingular bouncing cosmological models. In fact, it was found in \cite{Barragan:2009sq}  that the $f(R)$ model considered here admits homogeneous and isotropic bouncing and cyclic cosmologies in which the bounce occurs at a certain maximum energy density. The results presented here are compatible with such scenario, being the region between the apparent horizon and the bounce analogous to the contracting cosmological branch, while the expanding branch corresponds to the formation of a finite-size, exponentially-expanding baby universe connected with the outer universe via a throat (or umbilical cord)~\cite{Visser:1995cc}.

\begin{figure}[b]
    \centering
    \includegraphics[width=0.9\linewidth]{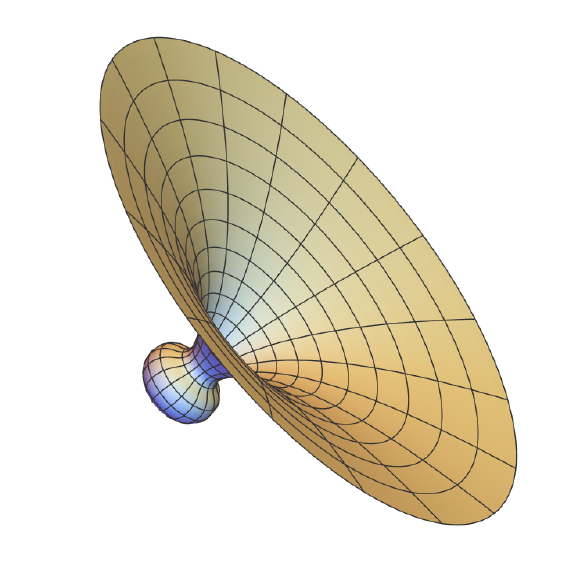}
    \caption{Embedding diagram of the spacetime geometry for the gravitational collapsing model B(p) in which the formation of a baby universe can be observed.
    }
    \label{fig:embedding}
\end{figure}

In Figure~\ref{fig:sizeuni}, we monitor the maximum reachable value of the areal radius within the baby universe region, $\tilde{R}^2_\text{max}$. Due to the singularity avoiding gauge chosen, we cannot observe regions close to the origin for large periods of time since they eventually extend beyond our computational grid. Nonetheless, we are able to follow the growth of the baby universe from $t\approx84.6$ to $t\approx 90.45$, observing that during this period of time the growth follows an exponential law.

A snapshot of the formation of this structure is depicted in Figure~\ref{fig:embedding}. Our simulations indicate that this cosmic bounce scenario is always hidden behind a horizon, hence causally disconnecting the baby universe from  observers above the horizon. A comprehensive analysis of this particular kind of evolution was recently reported in \cite{Maso-Ferrando:2023nju}, to which the interested reader is addressed for further details.

\section{Final remarks}\label{sec:final}
We have investigated the time evolution of spherically symmetric boson stars in Palatini $f(\R)$ gravity, focusing on the quadratic  model $f(\R)=\R+\xi\R^2$. We  compared the obtained solutions with those in GR, and explored both positive and negative values of the coupling parameter $\xi$. Our results reveal interesting differences when compared to GR models.

For  models A(n), A(z), and A(p), we obtain stable evolutions in which the parameters of stars, such as mass or shape, remain largely unchanged except for minor oscillations due to numerical errors coming from the discretization scheme. We note that in the case of model A(z) within the framework of GR, these oscillations are mainly attributed to the resolution of the initial data grid and would vanish in the continuum limit, as expected. However, for the  $f(\R)$ models A(n) and A(p), the change of coordinates, from polar-areal to isotropic, and the subsequent interpolation to accommodate the desired isotropic grid introduce a small source of numerical noise, which causes an artificial oscillation of the model that does not disappear with increasing resolution. This numerical error does not change the qualitative outcome of the evolution. Further research would be needed to obtain the initial data in an isotropic grid in order to get rid of the coordinates transformation 
and shed further light on the origin of those oscillations.

Our simulations have also shown that the unstable models B(n), B(z), and B(p) experience a migration towards the corresponding boson star configurations in the stable branch when perturbed only by discretization errors. However, when these three models were perturbed beyond the discretization error, they underwent gravitational collapse. In the context of GR, this leads to the formation of a black hole. In contrast, in the model B(p)  a richer internal structure emerges below the horizon due to the modified gravitational dynamics. In this case, a finite-size, exponentially-expanding baby universe connected with the outer universe via a throat was observed (see also~\cite{Maso-Ferrando:2023nju}), producing a scenario compatible with the notion of black bounce and nonsingular black holes proposed in recent literature \cite{Gambini:2020nsf,Simpson:2018tsi,Ashtekar:2023cod,Lobo:2020ffi}. Regarding the perturbed B(n) model, we have found that the approach used in this work is not suitable for computing its gravitational collapse fully, due to the appearance of divergences. This suggests the need for alternative approaches or refinements in computational techniques to properly analyze the gravitational collapse behavior of this specific model (or models within a theory with a negative value of the coupling parameter).

Finally, the unstable models C(n), C(z), and C(p), characterized by rapid decreases in $\Phi_0$, exhibited drastic radial expansion of the boson stars, ultimately resulting in their complete dispersion.

The study reported in this paper provides valuable insights into the dynamics and time evolution of boson stars in Palatini $f(\R)$ gravity, revealing notable differences compared to GR models. These differences emphasize the profound influence of the gravitational theory on the behavior and ultimate fate of self-gravitating compact objects like boson stars. Our findings open up avenues for further investigations and analyses, such as exploring gravitational collapse in other alternative gravity models, investigating additional features of boson stars (e.g.~adding the effects of rotation or self-interaction), or studying other types of compact objects (e.g.~Proca stars). By pursuing these avenues, we can deepen our understanding of the dynamics of exotic compact objects beyond the domain of GR.

\acknowledgements
AMF is supported by the Spanish Ministerio de Ciencia e Innovaci\'on with the PhD fellowship PRE2018-083802. NSG is supported by the Spanish Ministerio de Universidades, through a Mar\'ia Zambrano grant (ZA21-031) with reference UP2021-044, funded within the European Union-Next Generation EU.
This work is also supported by the Spanish Agencia Estatal de  Investigaci\'on (grants PID2020-116567GB-C21 and 
PID2021-125485NB-C21 funded by MCIN/AEI/10.13039/501100011033 and ERDF A way of making Europe) and by the Generalitat Valencia (Prometeo grants CIPROM/2022/49 and PROMETEO/2020/079). Further support is provided by the EU's Horizon 2020 research and innovation (RISE) programme H2020-MSCA-RISE-2017 (FunFiCO-777740) and  by  the  European Horizon  Europe  staff  exchange  (SE)  programme HORIZON-MSCA-2021-SE-01 (NewFunFiCO-101086251).  

\bibliography{bibliography}

\clearpage
\newpage
\appendix
\section{Convergence analysis}\label{sec:convergence}

The total mass of the spacetime can be calculated by integrating the stress-energy tensor at each spatial hypersurface $\Sigma$ \cite{Herdeiro:2016tmi}
\begin{equation}\label{eq:massa_kg}
    M = \int_{\Sigma} 
    \left(2T_t^t-T_\mu^\mu\right)\alpha\sqrt{\gamma} \,dr\,d\theta\, 
    d\varphi\ .
\end{equation}

\begin{figure*}[t]
\centering \includegraphics[width=0.47\linewidth]{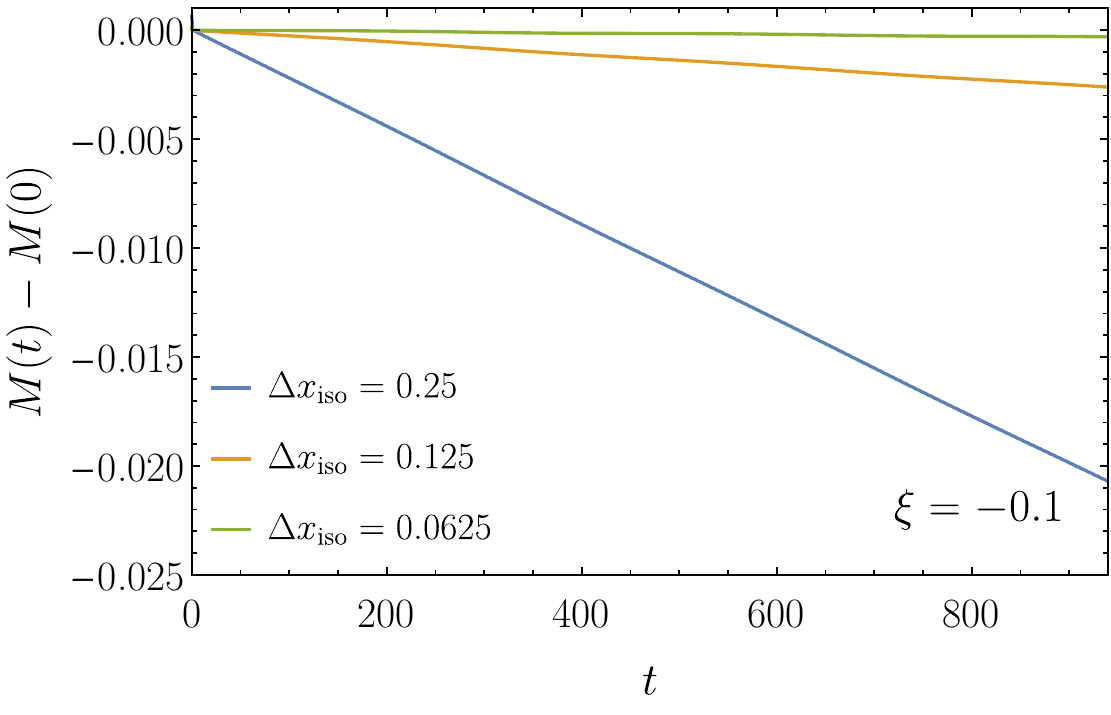}\hspace{5mm}\includegraphics[width=0.47\linewidth]{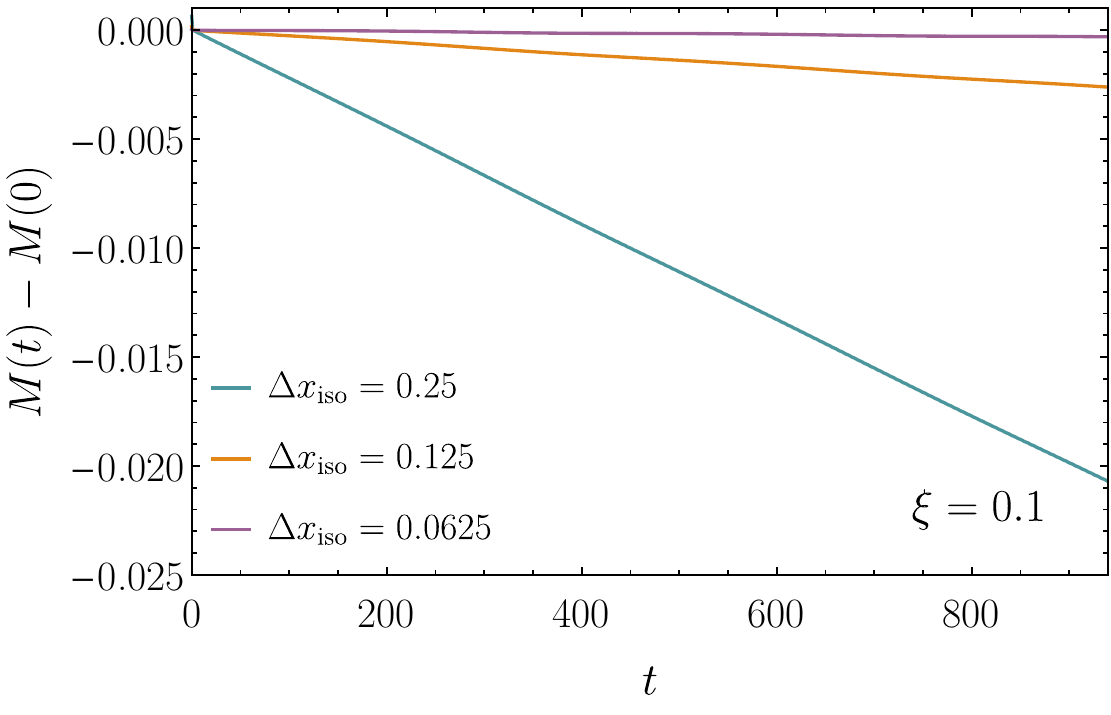}\\
\centering \includegraphics[width=0.47\linewidth]{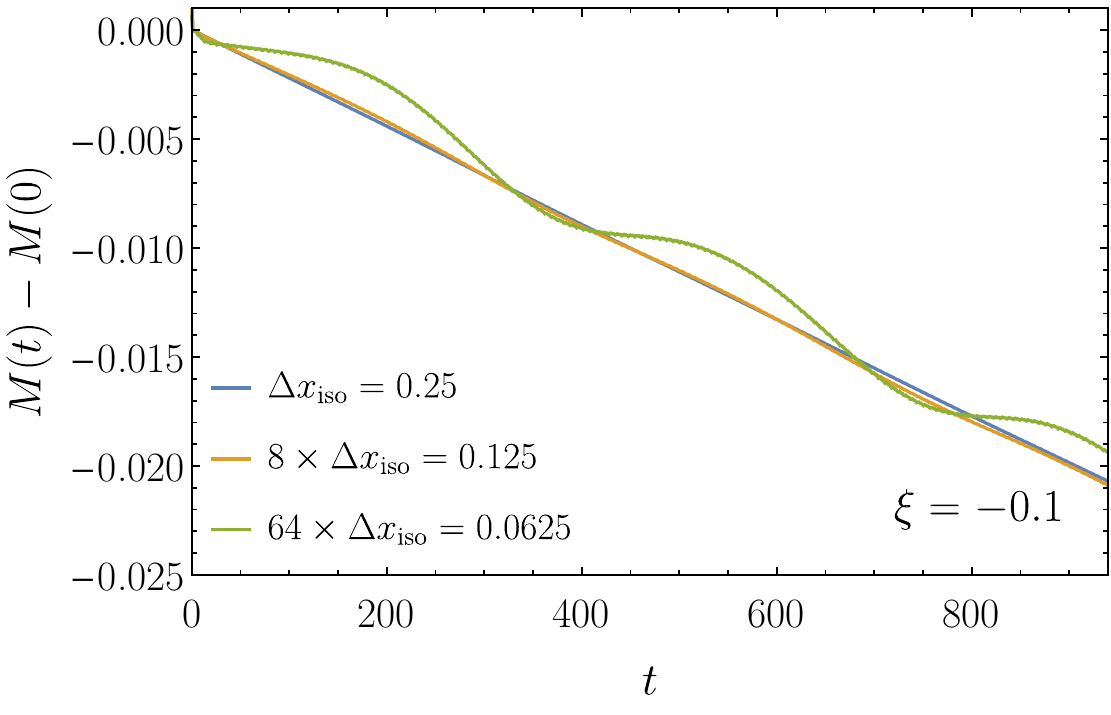}\hspace{5mm}\includegraphics[width=0.47\linewidth]{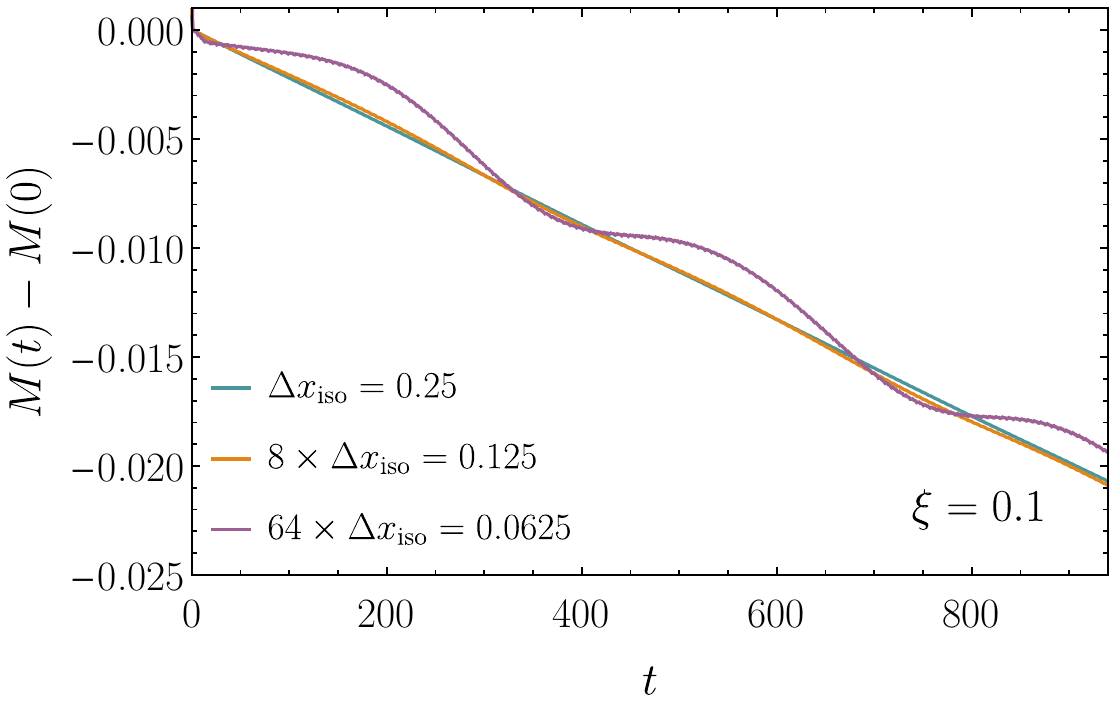}
\caption{Time evolution of the total mass of stable boson stars models A(n) (left panels) and A(p) (right panels). Top panels: Difference of the instantaneous total mass and its initial value for three different evolution grid resolutions $\Delta x$. Bottom panels: The quantities of the top panel are rescaled to show third-order convergence.}
\label{fig:stable}
\end{figure*}

In this analysis we are only considering the numerical error coming from the finite-differencing of the differential equations. This dominates the error if we use resolutions coarser than that used to compute the initial data. However, we note that the change of coordinates from polar-areal to isotropic (see details on the specific transformation in~\cite{Escorihuela-Tomas:2017uac}) also introduces an additional source of error.

Setting $\Delta x_{\text{pa}}=0.0025$ (the spatial resolution needed in the polar-areal grid used to compute the initial data) and choosing three resolutions for the isotropic grid, namely $\Delta x=0.25$, $\Delta x=0.125$ and $\Delta x=0.0625$, we find third-order convergence.

The results are plotted in Figure~\ref{fig:stable}. For the A(p) model, numerical errors from finite-differencing dominate the evolution and the total mass decreases with a drift that depends on resolution (see top panel of Fig.~\ref{fig:stable}). The rate of convergence of the total mass for this stable model is third order, as shown in the bottom panel of Fig.~\ref{fig:stable}.

The interested reader in the convergence analysis for the unstable models is addressed to \cite{Maso-Ferrando:2023nju}.

\end{document}